\documentclass[preprint,12pt]{elsarticle}

\usepackage[margin=2cm]{geometry}
\usepackage{algorithm}
\usepackage{textcomp}
\usepackage{stfloats}
\usepackage{url}
\usepackage{verbatim}

\usepackage{graphicx}%
\usepackage{amsthm}%
\usepackage{amssymb} 
\usepackage{amsmath}
\usepackage{mathrsfs}%
\usepackage{xcolor}%
\usepackage{textcomp}%
\usepackage{manyfoot}%
\usepackage{booktabs}%
\usepackage{algorithm}%
\usepackage{algorithmicx}%
\usepackage{algpseudocode}%
\usepackage{listings}%

\usepackage[hidelinks]{hyperref}
\usepackage{array}
\usepackage{subcaption}
\usepackage{caption}
\usepackage{multirow}
\usepackage{tabularx}

\newcolumntype{Y}{>{\centering\arraybackslash}X}

\usepackage{graphicx}
\usepackage{float}
\usepackage{fontawesome} 

\usepackage{amsmath,bm}
\usepackage{relsize} 
\usepackage{tikz}
\usetikzlibrary{shapes.multipart}
\usetikzlibrary{positioning}
\usetikzlibrary{shadows}
\usetikzlibrary{calc}

\usepackage{pgfplots} 
\pgfplotsset{compat=1.8}
\usepgfplotslibrary{statistics}
\tikzset{every picture/.style={/utils/exec={\ttfamily}}}
\tikzset{%
every picture/.style={/utils/exec={\sffamily}}
}

\def\ojoin{\setbox0=\hbox{$\bowtie$}%
  \rule[-.02ex]{.25em}{.4pt}\llap{\rule[\ht0]{.25em}{.4pt}}}
\def\leftouterjoin{\mathbin{\ojoin\mkern-5.8mu\bowtie}}

\usepackage{color, colortbl}
\definecolor{Gray}{gray}{0.925}
\definecolor{codegreen}{rgb}{0,0.6,0}
\definecolor{codegray}{rgb}{0.5,0.5,0.5}
\definecolor{codepurple}{rgb}{0.58,0,0.82}
\definecolor{backcolour}{rgb}{0.95,0.95,0.92}
\definecolor{BrickRed}{rgb}{0.8, 0.25, 0.33}
\definecolor{Dandelion}{rgb}{0.94, 0.88, 0.19}
\definecolor{OliveGreen}{rgb}{0.42, 0.56, 0.14}
\definecolor{light-gray}{gray}{0.85} 

\journal{}

\begin{document}

\begin{frontmatter}



\title{On the effects of logical database design on database size, query complexity, query performance, and energy consumption}


\author{Toni Taipalus}




\begin{abstract}
Database normalization theory is the basis for logical design of relational databases. Normalization reduces data redundancy and consequently eliminates potential data anomalies, while increasing the computational cost of read operations. Despite decades worth of applications of normalization theory, it still remains largely unclear to what extent normalization affects database size and efficiency. In this study, we study the effects of database normalization using the Internet Movie Database (IMDb) public dataset and PostgreSQL. The results indicate, rather intuitively, that \textit{(i)} database size on disk is reduced through normalization from 1NF to 2NF by 10\%, but not from 2NF to 4NF, \textit{(ii)} the number of tables and table rows in total increase monotonically from 1NF to 2NF to 4NF, and that \textit{(iii)} query complexity increases with further normalization. Surprisingly, however, the results also indicate that \textit{(iv)} normalization from 1NF to 2NF increases throughput by a factor of 4, and consequently, \textit{(v)} energy consumption per transaction reduces by 74\% with normalization from 1NF to 2NF. The results imply that the gains of normalization from 2NF to 4NF in terms of throughput and energy consumption are minimal, yet increase the storage space requirements by approximately 7\%. While these results represent merely one specific case, they provide needed empirical evaluation on the practical effects and magnitude of database normalization.
\end{abstract}




\end{frontmatter}


\section{Introduction}
\label{sec-introduction}

Database normalization has existed since the dawn of relational databases \cite{Codd_1970}, serving as the fundamental design principle behind logical database design. While it has been argued that normalization is justified due to normalized data structures being mathematically \cite{Date_1986_relational} and cost-efficiently \cite{Westland_1992} simpler, normalized databases are arguably \textit{intuitively} more complex, since they contain more tables than non-normalized databases. The notion that the highest normal form is not always the most feasible one is commonly accepted, and relational databases are seldom normalized to the highest normal form \cite[][pp. 361-363]{Date_1986_relational}\cite{Albarak_2018_prioritizing}. The general reluctance to always normalize to the highest normal form is perhaps due to this increase in intuitive complexity, normalization being accepted as a challenging task \cite{Date_2019_all_that_jazz}\cite{Fischer_2023}, or that normalization is, especially with an already deployed database, relatively arduous \cite{Albarak_2020}. Some studies have concluded that it is not a trivial task to decide which normal form should be the target of any given business domain \cite{Lee_1995}.

In the last decade or so, software industry and research has been paying more and more attention to the energy consumption of software \cite{Cruz_2019} and software infrastructure as a whole \cite{Chien_2019,Knowles_2022}. Database normalization is a relatively cost-efficient way of of potentially reducing energy consumption of transaction processing systems, as a system that can serve more customers with the same hardware and same energy consumption is arguably desirable. Furthermore, database normalization has the intuitive potential of reducing the carbon footprint resulting from storage hardware, as database normalization strives towards minimizing data redundancy. Although the effects of database normalization are widely accepted, the magnitude of these effects has not received much scientific attention, possibly due to the fact that fair performance testing is considered challenging \cite{Manolescu_2009,Taipalus_2024_perf}. To this end, we chose the Internet Movie Database (IMDb) with real-world data and PostgreSQL to test the implications of database normalization on database size, query complexity, query performance, and energy consumption. We have strived to follow performance comparison guidelines reported in scientific literature, to report our testing transparently, and to emphasize that our findings may be limited to this specific use case.

Our results show that in the case of IMDb and secondary indices designed for our test suite of seven SQL queries, database size on disk decreased by 10.3\% when the database was normalized from first normal form (1NF) to 2NF, yet increased by 6.5\% with normalization from 2NF to 4NF. Normal forms between 2NF and 4NF were not considered, as the database structure was the same with 3NF, BCNF and 4NF. The number of tables increased from 8 (1NF) to 10 (2NF) to 13 (4NF). Query complexity in terms of number of tables and table joins in queries either increased or remained the same, depending on the query. With a single-node, read-only workload with 48 concurrent clients, throughput increased by 311\% when the database was normalized from 1NF to 2NF, and decreased by less than 1\% when the database was normalized from 2NF to 4NF. Because of the differences in throughput, yet relatively similar power consumption profiles, the transaction processing using the 1NF database was 73.9\% less energy efficient when compared to the 2NF database, meaning that processing one transaction in the 2NF database consumed approximately fourth of the energy needed for a transaction in the 1NF database. Differences between the 2NF and 4NF databases were negligible.

The rest of the paper is structured as follows. In the next section, we discuss related terminology and related prior works on database normalization and performance. In Section~\ref{sec-setting} we describe the IMDb database, our normalization process, the test suite, and PostgreSQL configuration. Section~\ref{sec-results} contains the results. In Section~\ref{sec-discussion} we discuss the practical implications, threats to validity, and future research avenues. Instead of taking the results presented herein as is, we urge the reader to consider the potential pitfalls and limitations of the study which are discussed in Section~\ref{sec-disc-threats}. Section~\ref{sec-concl} concludes the study.


\section{Background}
\label{sec-bg}

\subsection{Database normalization}
\label{sec-bg-nfs}

In the subsequent sections, we use the terms \textit{database}, \textit{schema} and \textit{database schema} interchangeably. Additionally, and while we acknowledge the theoretical and practical differences associated with terms in relational theory and terms in SQL, we use the terms \textit{table} and \textit{relation}, and \textit{column} and \textit{attribute} interchangeably for brevity. Some terms are a better fit for some contexts, e.g., attributes and relations in normalization, and columns and tables in their respective implementations in SQL.

There are many approaches to relational database design, all of which largely rely on individual intuition regarding table and column names, and the placement of specific columns to specific tables. Normal forms may be considered a formal approach to minimizing the effects of human intuition on relational database design, with the objective of minimizing data redundancy and consequent, potential data anomalies \cite{Atzeni_1982}. Normalization can be considered an iterative process, starting from low normal forms and proceeding to higher normal forms until the target normal form is reached. In each iteration, tables that violate the target normal form are typically divided vertically into smaller tables. Each iteration potentially (but not necessarily) results in a higher number of tables in the database. Several algorithms have been proposed for the normalization process to be automated to certain degrees \cite{Link_2021}.

Database normalization is largely concerned with finding \textit{functional dependencies} \cite{Codd_1970} between the attributes of a given relation. In layperson's terms, a functional dependency means that the value of a set of attributes determines the value of another attribute, e.g., the value of a car's license plate determines the make and model of the car, but not vice versa. There are also several special cases of functional dependencies such as \textit{trivial} functional dependencies which cannot be violated, \textit{multivalued} functional dependencies, by which the value of a set of attributes determines a set of values of another attribute, and \textit{join} dependencies, which are concerned with certain combinations of data from different relations that can be combined to form all the data in another relation.

A relational database table which adheres to the requirements defined by the relational theory is said to adhere to first normal form (1NF) \cite{Codd_1970}. SQL implementations do not require that tables adhere to the criteria determined by the relational theory, and it is possible that tables do not have a set of attributes which differentiate the rows in the table (i.e., candidate keys), or that tables may contain non-atomic columns. From a database normalization perspective, we call such tables non-first normal form tables (NFNF). The second normal form (2NF) dictates that table columns are required to be fully functionally dependent on a candidate key of the table. The third normal form (3NF), in turn, prohibits transitional functional dependencies on the table's candidate key, i.e., a dependencies which hold through a chain of dependencies. Furthermore, the Boyce/Codd normal form (BCNF) dictates that each functional dependency in a relation much be determined by a candidate key (or a superset of one). Finally, the fourth normal form (4NF) requires that in each non-trivial multivalued functional dependency is determined by a candidate key (or a superset of one). The aforementioned definitions typically implicitly subsume that the requirements of the previous normal form are met. There are several other normal form in addition to these, but they are not relevant to this work \cite[cf. e.g.,][]{Kohler_2016}.

As stated previously, normalization can decrease the need for storage space by eliminating redundant data. Typically, each normal form, being more demanding than the previous, imposes requirements on the relations. When the requirements of a target normal form is not met, a relation is typically divided vertically, resulting in a more complex database structure, but less data. How much the size of the dataset is reduced typically depends on the cardinality of values. For example, consider table \texttt{A(a\_id, type, country)} which holds 1,000 rows of data. \texttt{a\_id} is a 4 byte integer, acting as an identifier, \texttt{type} is a 4 byte integer with arbitrary values ranging from 1 to 500, and \texttt{country} is a 15 byte string with 50 different values. Functional dependencies \{a\_id\} $\rightarrow$ \{type\} and \{type\} $\rightarrow$ \{country\} hold, meaning that the table is in violation of 3NF and the structure contains redundancy. Simplified, the table requires 24,000 bytes of storage space (1,000 $\times$ 4 + 1,000 $\times$ 4 + 1,000 $\times$ 16). If the table is divided into tables \texttt{A\textsubscript{1}(a\_id, type)} and \texttt{A\textsubscript{2}(type, country)} to satisfy 3NF, the storage space requirement is reduced to 18,000 bytes (1,000 $\times$ 4 + 1,000 $\times$ 4 + 500 $\times$ 4 + 500 $\times$ 16), a reduction of 25\%. However, if the cardinality of values in the original table were different, the effects of normalization can be smaller or greater. If there were merely 5 different values of \texttt{type}, the storage requirement for the normalized tables would be 8,100 bytes (1,000 $\times$ 4 + 1,000 $\times$ 4 + 5 $\times$ 4 + 5 $\times$ 16), increasing the reduction to 66.25\%. Due to less data redundancy, different write-operation induced data anomalies are potentially less common.

Despite the fact that normalization potentially reduces data anomalies and storage space requirements, the consequent more complex data structures pose potential decreases in performance. In the aforementioned database of one table (\texttt{A}), querying all information from the database requires a query with no table joins, while in the normalized version of the database (tables \texttt{A\textsubscript{1}} and \texttt{A\textsubscript{2}}), the query requires one table join. Joins are typically considered one of the most computationally costly query-related operations in a relational database. In theory, this means that read operations in a more normalized database are computationally more costly than queries in a less normalized database. On the other hand, database normalization may decrease the computational cost of certain write operations. For example, if the table \texttt{A}, where there are 5 different values of \texttt{type}, and consequently, 5 different values of \texttt{country}, updating the name of one country requires updating as many as 200 values on average (1,000 rows $\div$ 5 different values). In contrast, updating the name of one country in the normalized database with tables \texttt{A\textsubscript{1}} and \texttt{A\textsubscript{2}} requires that only one value is updated.

\subsection{Facets of database efficiency}
\label{sec-bg-efficiency}

There are several facets to database efficiency. Perhaps the most intuitive efficiency metrics are transaction latency and throughput. Especially latency is crucial for the end-user, as systems which respond fast are naturally more pleasant to use \cite{Loukopoulos_2002}. In contrast, high throughput is especially important for the service provider, as it is crucial that the system is able to cater for the needs of many end-users simultaneously, and preferable utilizing hardware such as memory and CPU cost-efficiently. There are several benchmarks developed by the Transaction Processing Council (TPC) which measure the throughput of concurrent transactions in different environments such as banks (TPC-A) and warehouses (TPC-C), as well as several others such as OLTP-Bench \cite{Difallah_2013}, JOB \cite{Leis_2015}, and PeakBench \cite{Zhang_2020_peakbench}. These benchmarks can give estimations on how performance is affected by aspects such as different hardware and software configurations, DBMS parameters, and dataset sizes, yet is has been questioned whether current benchmarking software are adequate for more complex environments such as distributed databases \cite{Qu_2022}.

Efficiency also may be measured with other metrics such as storage space requirements and energy consumption. As it is common that databases are replicated across multiple computing nodes to facilitate both performance and fault tolerance, storage space requirements arguably play a role significant enough to warrant attention \cite{Shi_2020}. Besides fault tolerance considerations, one justification for database replication may be to bring the database geographically closer to the end-user, thus reducing network latency. Another reason may be to balance the load of end-users across different computing nodes, each with a local replica of the database. For example, if a database with a size of 100 GB is fully replicated across 12 computing nodes, a storage space increase of 10\% rises from 10 GB to 120 GB, which potentially requires more storage hardware.

Finally, computing consumes resources not limited to hardware. According to one estimate, computing will account for 21\% of the world's energy consumption by 2030 \cite{Jones_2018}, and this 2018 estimate probably did not foresee the effects of the rapid popularisation of artificial intelligence services in early 2020s. Estimations such as this are probably one of the reasons data systems research has focused more and more on green data or data system energy consumption \cite{Wu_2016,Guo_2022,Beloglazov_2011}, as data is one of the major parts of any software system. The challenge in DBMS energy consumption is that query optimizers are typically cost-based with the primary objective of executing queries as fast as possible, energy consumption being a secondary concern at best \cite{Guo_2017}. Energy consumption ($E$, in Joules) of workload $w$ over time interval $[0,T]$ can be calculated with the integral $E(w,T) = \int_{0}^{T} P(w, t) \, dt$, which sums up the instantaneous power ($P$, in watts) of the signal over each infinitesimal increment of time within that interval \cite[cf. e.g.,][]{Guo_2022}. 

\subsection{Related work}
\label{sec-bg-relworks}

Despite the age of relational database theory and practice, research on the the effects of database normalization on database efficiency has remained in the sidelines. However, some scholars have studied the effects of database normalization, and many scholars have studied the performance of DBMSs. On one hand, outside the realm of database performance, studies have shown that querying normalized data structures are more error-prone \cite{Bowen_2009,Borthick_2001}, and this is most likely due to the increase in the complexity of the database structures \cite{Taipalus_2020_effects}, i.e., more tables and columns. From a more holistic perspective, some studies have argued that tables not normalized up to 4NF potentially manifest as technical debt \cite{Albarak_2020,Albarak_2018_identifying}, and that database normalization should be guided by a systematic process, as the highest normal forms are rarely the most suitable ones \cite{Albarak_2018_prioritizing}. As database structure refactoring, especially in relational DBMSs, is a relatively arduous process, paying the debt through normalization is significantly more costly further down the development process.

On the other hand, in the realm of database performance but outside the realm of normalization, a recent systematic literature review \cite{Taipalus_2024_perf} observed over a hundred scientific studies comparing the performance of one DBMS to another, concluding that most of these studies \textit{(i)} do not test performance with use-cases that try to mimic real-world situations, \textit{(ii)} are not reported transparently enough for replication, and \textit{(iii)} do not follow documented (and sometimes, rather intuitive) guidelines for performance testing \cite[cf. e.g.,][]{Raasveldt_2018,Gray_1992,Dietrich_1992}. Furthermore, measuring DBMS performance is prone to errors due to the multitude and complex nature of different parameters and their, sometimes unpredictable, interdependencies \cite{Taipalus_2024_perf}. Some of these parameters are related to hardware, some to DBMS settings and features, some to database structures such as tables and indices, some to how queries are formulated, and some to concurrency and database distribution.

There are some closely related studies which examined the effects of normalization on performance. One empirical study found that \textit{de}normalization (e.g., transforming tables from 3NF to 2NF instead of from 2NF to 3NF), positively affected query performance as much as 15\% \cite{Chun_2019}. In contrast, some studies have argued for the performance benefits of denormalization by presenting general (as opposed to domain-specific) methods and guidelines for determining when to normalize for performance gains \cite{Sanders_2001,Westland_1992,Lee_1995}. It is unclear if these methods have been tested in practice and how they have been validated in real-world reflecting scenarios. Overall, it is surprising how little the effects of database normalization on performance has been empirically studied, given that there are over a hundred studies in the past 15 years on DBMS performance comparisons.


\section{Research setting} 
\label{sec-setting}

\subsection{Data preparation}


We started the data preparation by downloading the IMDb public dataset\footnote{https://datasets.imdbws.com/} and importing the data into a PostgreSQL database. The dataset is constantly updated, and ours was downloaded on December 2nd, 2023. At that time, the IMDb database contained 9,815,519 titles (i.e., movies, TV-series episodes, etc.) with associated actors and their roles, and other associated crew such as directors and writers. The public dataset is not complete in either database structure nor data. For example, movie reviews, which arguably represent the most write-heavy aspect of the database, are missing from the dataset. No primary or foreign keys were defined, as they are of little concern in read-only databases such as the one in our tests.

\subsection{Denormalization and normalization}
\label{sec-setting-normalization}

\begin{figure*}
\centering
    \begin{subfigure}[b]{0.49\textwidth}
    \centering
\tikzset{%
    mylabel/.style={font=\normalsize},
    pics/entity/.style n args={3}{code={%
        \node[draw,
        rectangle split,
        rectangle split parts=2,
        text height=1.5ex,
        text width=9em,
        ] (#1)
        {#2 \nodepart[font=\small]{second}
            \begin{tabular}{>{\raggedright\arraybackslash}p{12em}}
                #3
            \end{tabular}
        };%
    }}
}
\resizebox{0.98\linewidth}{!}{
\begin{tikzpicture}
    \pic {entity={tb}{title\_basics}{%
        tconst \\
        titletype \\
        primarytitle \\
        originaltitle \\
        isadult \\
        startyear \\
        endyear \\
        runtimeminutes \\
        genres[]
    }};
    \pic[left=3em of tb] {entity={ta}{title\_akas}{%
        titleid \\
        ordering \\
        title \\
        region \\
        language \\
        types[] \\
        attributes[] \\
        isoriginaltitle
    }};
    \pic[below=1em of ta] {entity={tp}{title\_principals}{%
        tconst \\
        ordering \\
        nconst \\
        category \\
        job \\
        characters[]
    }};
    \pic[below=1em of tb] {entity={nb}{name\_basics}{%
        nconst \\
        primaryname \\
        birthyear \\
        deathyear \\
        primaryprofession[] \\
        knownfortitles[]
    }};
    \pic[right=3em of tb] {entity={te}{title\_episodes}{%
        tconst \\
        parenttconst \\
        seasonnumber \\
        episodenumber
    }};
    \pic[below=1em of te] {entity={tc}{title\_crew}{%
        tconst \\
        directors[] \\
        writers []
    }};
    \pic[below=1em of tc] {entity={tr}{title\_ratings}{%
        tconst \\
        averagerating \\
        numvotes
    }};
    
    \draw[] (ta.east) -- (tb.west);
    \draw[] ([yshift=2.75]tp.east) --++ (.5,0) |- ([yshift=-5]tb.west);
    \draw[] ([yshift=-4.5]tp.east) -- (nb.west);

    \draw[] (te.west) -- (tb.east);
    \draw[] ([yshift=0]tc.west) --++ (-0.4,0) |- ([yshift=-5]tb.east);
    \draw[] ([yshift=0]tr.west) --++ (-0.6,0) |- ([yshift=-10]tb.east);
    
\end{tikzpicture}
}
    \vspace{0\baselineskip}
    \caption{Baseline schema (NFNF)}
    \label{fig-schema-baseline}
\end{subfigure}
\begin{subfigure}[b]{0.49\textwidth}  
    \centering 
\tikzset{%
    mylabel/.style={font=\normalsize},
    pics/entity/.style n args={3}{code={%
        \node[draw,
        rectangle split,
        rectangle split parts=2,
        text height=1.5ex,
        text width=11em,
        ] (#1)
        {#2 \nodepart[font=\small]{second}
            \begin{tabular}{>{\raggedright\arraybackslash}p{12em}}
                #3
            \end{tabular}
        };%
    }}
}
\resizebox{0.98\linewidth}{!}{
\begin{tikzpicture}
    \pic {entity={tb}{title\_basics}{%
        tconst \\
        ordering \\
        title \\
        titletype \\        
        isoriginaltitle \\
        region \\
        language \\
        isadult \\
        startyear \\
        endyear \\
        runtimeminutes \\
        averagerating \\
        numvotes 
    }};
    \pic[left=3em of tb] {entity={tg}{title\_genres}{%
        tconst \\
        genre
    }};
    \pic[below=1em of tg] {entity={tdw}{title\_directors\_writers}{%
        tconst \\
        director \\
        writer 
    }};
    \pic[below=1em of tdw] {entity={tc}{title\_characters}{%
        tconst \\
        nconst \\
        character
    }};
    \pic[below=1em of tb] {entity={nb}{name\_basics}{%
        nconst \\
        tconst \\
        ordering \\
        category \\
        job \\
        primaryname \\
        birthyear \\
        deathyear
    }};
    \pic[right=3em of tb] {entity={te}{title\_episodes}{%
        tconst \\
        tconst\_parent \\
        seasonnumber \\
        episodenumber
    }};
    \pic[below=1em of te] {entity={tya}{title\_types\_attributes}{%
        tconst \\
        type \\
        attribute
    }};
    \pic[below=1em of tya] {entity={npt}{name\_professions\_titles}{%
        nconst \\
        primaryprofession \\
        knownfortitle
    }};
    
    \draw[] (tg.east) -- (tb.west);
    \draw[] ([yshift=2.75]tdw.east) --++ (.3,0) |- ([yshift=-5]tb.west);
    \draw[] ([yshift=2.75]tdw.west) --++ (-.3,0) |- ([yshift=-20]nb.west); 
    \draw[] ([yshift=2.75]tc.east) --++ (.7,0) |- ([yshift=-10]tb.west);
    \draw[] ([yshift=-3]tc.east) --++ (.7,0) |- ([yshift=-0]nb.west);

    \draw[] (te.west) -- ([yshift=0]tb.east);
    \draw[] ([yshift=0]tya.west) --++ (-0.5,0) |- ([yshift=-5]tb.east);
    \draw[] ([yshift=5]npt.west) --++ (-0.7,0) |- ([yshift=-30]tb.east);
    \draw[] (npt.west) -- ([yshift=7]nb.east);

    \draw[] (nb.north) -- (tb.south);
    
\end{tikzpicture}
}
    \vspace{0\baselineskip}
    \caption{1NF schema}  
    \label{fig-schema-1nf}
\end{subfigure}
\vskip\baselineskip 
\begin{subfigure}[b]{0.49\textwidth}   
    \centering 
\tikzset{%
    mylabel/.style={font=\normalsize},
    pics/entity/.style n args={3}{code={%
        \node[draw,
        rectangle split,
        rectangle split parts=2,
        text height=1.5ex,
        text width=11em,
        ] (#1)
        {#2 \nodepart[font=\small]{second}
            \begin{tabular}{>{\raggedright\arraybackslash}p{12em}}
                #3
            \end{tabular}
        };%
    }}
}
\resizebox{0.98\linewidth}{!}{
\begin{tikzpicture}
    \pic {entity={tb}{title\_basics}{%
        tconst \\
        titletype \\        
        isadult \\
        startyear \\
        endyear \\
        runtimeminutes \\
        averagerating \\
        numvotes 
    }};
    \pic[left=3em of tb] {entity={tg}{title\_genres}{%
        tconst \\
        genre
    }};
    \pic[below=1em of tg] {entity={tdw}{title\_directors\_writers}{%
        tconst \\
        director \\
        writer 
    }};
    \pic[below=1em of tdw] {entity={tc}{title\_characters}{%
        tconst \\
        nconst \\
        character
    }};
    \pic[below=1em of tb] {entity={nb}{name\_basics}{%
        nconst \\
        primaryname \\
        birthyear \\
        deathyear
    }};
    \pic[above right=3em of tb] {entity={te}{title\_episodes}{%
        tconst \\
        tconst\_parent \\
        seasonnumber \\
        episodenumber
    }};
    \pic[below=1em of te] {entity={tak}{title\_akas}{%
        tconst \\
        ordering \\
        title \\
        isoriginaltitle \\
        region \\
        language
    }};
    \pic[below=1em of tak] {entity={tp}{title\_principals}{%
        nconst \\
        tconst \\
        ordering \\
        category \\
        job
    }};
    \pic[below=1em of tp] {entity={npt}{name\_professions\_titles}{%
        nconst \\
        primaryprofession \\
        knownfortitle
    }};
    \pic[above=1em of te] {entity={tya}{title\_types\_attributes}{%
        tconst \\
        type \\
        attribute
    }};
    
    \draw[] (tg.east) -- (tb.west);
    \draw[] ([yshift=2.75]tdw.east) --++ (.3,0) |- ([yshift=-5]tb.west);
    \draw[blue] ([yshift=-10]tdw.east) --++ (.3,0) |- ([yshift=10]nb.west); 
    \draw[] ([yshift=2.75]tc.east) --++ (.7,0) |- ([yshift=-15]tb.west);
    \draw[] ([yshift=-3]tc.east) --++ (.9,0) |- ([yshift=-0]nb.west);

    \draw[] (tak.west) -- ([yshift=22.5]tb.east);
    \draw[] ([yshift=0]te.west) --++ (-0.3,0) |- ([yshift=25]tb.east);
    \draw[] ([yshift=0]tp.west) --++ (-0.5,0) |- ([yshift=-10]tb.east);
    \draw[] ([yshift=-5]tp.west) --++ (-0.5,0) |- ([yshift=10]nb.east);
    \draw[] ([yshift=0]npt.west) --++ (-0.5,0) |- ([yshift=-5]nb.east);
    \draw[blue] ([yshift=5]npt.west) --++ (-0.3,0) |- ([yshift=-5]tb.east); 
    
    \draw[] ([yshift=0]tya.east) --++ (0.5,0) |- ([yshift=0]tak.east);
    
\end{tikzpicture}
}
    \vspace{0\baselineskip}
    \caption{2NF schema}
    \label{fig-schema-2nf}
\end{subfigure}
\begin{subfigure}[b]{0.49\textwidth}   
    \centering 
\tikzset{%
    mylabel/.style={font=\normalsize},
    pics/entity/.style n args={3}{code={%
        \node[draw,
        rectangle split,
        rectangle split parts=2,
        text height=1.5ex,
        text width=8.25em,
        ] (#1)
        {#2 \nodepart[font=\small]{second}
            \begin{tabular}{>{\raggedright\arraybackslash}p{12em}}
                #3
            \end{tabular}
        };%
    }}
}
\resizebox{0.98\linewidth}{!}{
\begin{tikzpicture}
    \pic {entity={tb}{title\_basics}{%
        tconst \\
        titletype \\        
        isadult \\
        startyear \\
        endyear \\
        runtimeminutes \\
        averagerating \\
        numvotes 
    }};
    \pic[left=3em of tb] {entity={tg}{title\_genres}{%
        tconst \\
        genre
    }};
    \pic[below=1em of tg] {entity={td}{title\_directors}{%
        tconst \\
        director
    }};
    \pic[below=1em of td] {entity={tw}{title\_writers}{%
        tconst \\
        writer
    }};
    \pic[below=1em of tw] {entity={tc}{title\_characters}{%
        tconst \\
        nconst \\
        character
    }};
    \pic[below=1em of tb] {entity={nb}{name\_basics}{%
        nconst \\
        primaryname \\
        birthyear \\
        deathyear
    }};
    \pic[above right=3em of tb] {entity={te}{title\_episodes}{%
        tconst \\
        tconst\_parent \\
        seasonnumber \\
        episodenumber
    }};
    \pic[below=1em of te] {entity={tak}{title\_akas}{%
        tconst \\
        ordering \\
        title \\
        isoriginaltitle \\
        region \\
        language
    }};
    \pic[below=1em of tak] {entity={tp}{title\_principals}{%
        nconst \\
        tconst \\
        ordering \\
        category \\
        job
    }};
    \pic[below=1em of tp] {entity={np}{name\_professions}{%
        nconst \\
        primaryprofession
    }};
    \pic[below=1em of np] {entity={nt}{name\_titles}{%
        nconst \\
        knownfortitle
    }};
    \pic[right=2em of tak] {entity={tat}{title\_attributes}{%
        tconst \\
        attribute
    }};
    \pic[below=1em of tat] {entity={ty}{title\_types}{%
        tconst \\
        type
    }};
    
    \draw[] (tg.east) -- (tb.west);
    \draw[] ([yshift=2.75]td.east) --++ (.3,0) |- ([yshift=-5]tb.west);
    \draw[blue] ([yshift=0]td.east) --++ (.3,0) |- ([yshift=15]nb.west); 
    \draw[] ([yshift=2.75]tw.east) --++ (.5,0) |- ([yshift=-10]tb.west);
    \draw[blue] ([yshift=17]tw.east) -- ([yshift=5.5]nb.west); 
    \draw[] ([yshift=2.75]tc.east) --++ (.7,0) |- ([yshift=-15]tb.west);
    \draw[] ([yshift=-3]tc.east) --++ (.9,0) |- ([yshift=-0]nb.west);

    \draw[] (tak.west) -- ([yshift=22.5]tb.east);
    \draw[] ([yshift=0]te.west) --++ (-0.3,0) |- ([yshift=25]tb.east);
    \draw[] ([yshift=0]tp.west) --++ (-0.5,0) |- ([yshift=-10]tb.east);
    \draw[] ([yshift=-5]tp.west) --++ (-0.5,0) |- ([yshift=10]nb.east);
    \draw[] ([yshift=0]np.west) --++ (-0.3,0) |- ([yshift=-5]nb.east);
    \draw[] ([yshift=0]nt.west) --++ (-0.5,0) |- ([yshift=-10]nb.east);
    \draw[blue] ([yshift=5]nt.west) --++ (-0.15,0) |- ([yshift=-5]tb.east); 

    \draw[] (tat.west) -- (tak.east);
    \draw[] ([yshift=0]ty.west) --++ (-0.5,0) |- ([yshift=-5]tak.east);

\end{tikzpicture}
}
    \vspace{0\baselineskip}
    \caption{4NF schema}  
    \label{fig-schema-4nf}
\end{subfigure}
    \caption{IMDb database schemas normalized into 1NF, 2NF and 4NF; square brackets after column names in the baseline schema represent compound values; some foreign keys -- although not enforced by implemented constraints -- are presented in blue to clarify overlapping foreign keys}
    \label{fig-schemas}
\end{figure*}
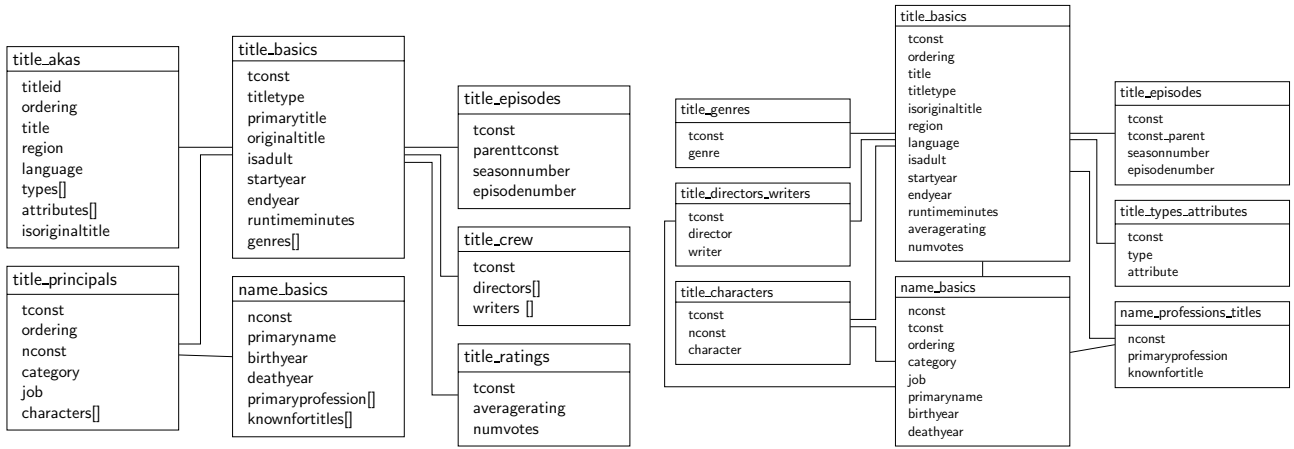
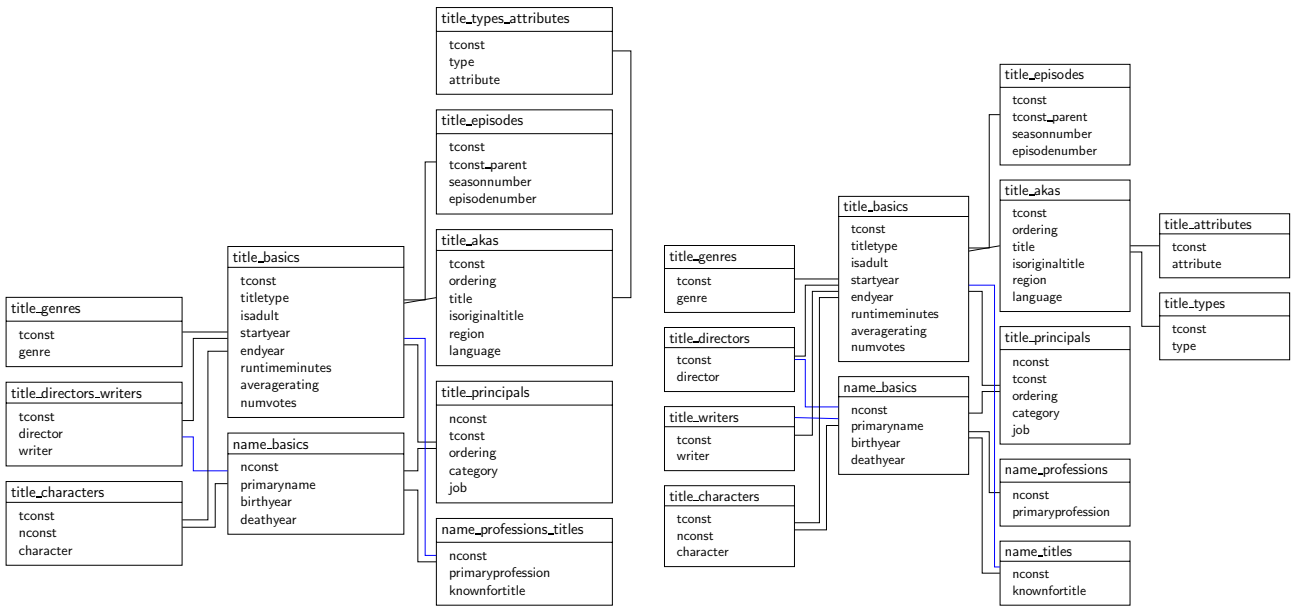

The original IMDb database (i.e., \textit{baseline} schema) contains compound attributes in the form of non-uniformly-defined lists stored as text (cf. Fig.~\ref{fig-schema-baseline}). Effectively, this means that the original database does not adhere to 1NF by design. We started the denormalization process by reducing the baseline schema to a group of attributes without host tables, and by unnesting the compound attributes. Next, we determined functional dependencies among all attributes, and formulated tables adhering to 1NF (cf. Fig.~\ref{fig-schema-1nf}) with at least one candidate key, effectively eliminating the possibility of an ``universal relation'' \cite{Atzeni_1982}. It is worth noting that the attribute \textit{region} in relation \textit{title\_basics} does not functionally determine attribute \textit{language} or vice versa, as several languages may be spoken in one region, and one language may be spoken in several regions. We did not define primary or foreign keys for any of the tables in any of the schemas, as we will measure only read performance in this study.

Next, we normalized a copy of the 1NF database schema further into 2NF. In the relation \textit{title\_basics} hold 2NF-violating functional dependencies \{tconst\} $\rightarrow$ \{titletype, isadult, startyear, endyear, runtimeminutes, averagerating, numvotes\}. This relation was divided into \textit{title\_basics} and \textit{title\_akas}. Additionally, in the relation \textit{name\_basics} hold 2NF-violating functional dependencies \{nconst\} $\rightarrow$ \{primaryname, birthyear, deathyear\}. This relation was divided into \textit{name\_basics} and \textit{title\_principals} (cf. Fig.~\ref{fig-schema-2nf}).

Next, we normalized a copy of the 2NF database schema further. We found no functional dependencies violating 3NF or BCNF. However, in relation \textit{title\_types\_attributes} hold 4NF-violating non-trivial multi-valued dependencies \{tconst\} $\twoheadrightarrow$ \{type, attribute\}, in relation \textit{title\_directors\_writers} \{tconst\} $\twoheadrightarrow$ \{director, writer\} and in relation \textit{name\_professions\_titles} \{nconst\} $\twoheadrightarrow$ \{primaryprofession, knownfortitle\}. These relations were divided (cf. Fig.~\ref{fig-schema-4nf}). Although the 2NF schema adheres to BCNF, we chose to call it 2NF, as that was the target normal form. Furthermore, our 4NF may adhere to normal forms higher than 4NF, but those normal forms were not considered.

\begin{table*}
\caption{Database schemas normalized from the the IMDb baseline database; abbreviations are used in the query trees in Section~\ref{sec-results-query-complexity}; all tables have only one candidate key}
\label{table-schemas}
\centering
\begin{tabular}{clll}
\hline
  &\multicolumn{1}{l}{Table name} &\multicolumn{1}{l}{Abbr.} & \multicolumn{1}{l}{Candidate key} \\
 \hline
 \parbox[t]{2mm}{\multirow{8}{*}{\rotatebox[origin=c]{90}{1NF}}} 
 & name\_basics              & nb  & \{nconst, tconst, ordering\}                 \\
 & name\_professions\_titles & npt & \{nconst, primaryprofession, knownfortitle\} \\
 & title\_basics             & tb  & \{tconst, ordering\}                         \\
 & title\_characters         & tc  & \{tconst, nconst, character\}                \\
 & title\_directors\_writers & tdw & \{tconst, director, writer\}                 \\
 & title\_episodes           & te  & \{tconst\}                                   \\
 & title\_genres             & tg  & \{tconst, genre\}                            \\
 & title\_types\_attributes  & tya & \{tconst, type, attribute\}                  \\
 \hline
 \parbox[t]{2mm}{\multirow{10}{*}{\rotatebox[origin=c]{90}{2NF}}} 
 & name\_basics              & nb  & \{nconst\}                                   \\
 & name\_professions\_titles & npt & \{nconst, primaryprofession, knownfortitle\} \\
 & title\_akas               & ta  & \{tconst, ordering\}                         \\
 & title\_basics             & tb  & \{tconst\}                                   \\
 & title\_characters         & tc  & \{tconst, nconst, character\}                \\
 & title\_directors\_writers & tdw & \{tconst, director, writer\}                 \\
 & title\_episodes           & te  & \{tconst\}                                   \\
 & title\_genres             & tg  & \{tconst, genre\}                            \\
 & title\_principals         & tp  & \{nconst, tconst, ordering\}                 \\
 & title\_types\_attributes  & tya & \{tconst, type, attribute\}                  \\
 \hline
 \parbox[t]{2mm}{\multirow{13}{*}{\rotatebox[origin=c]{90}{4NF}}} 
 & name\_basics      & nb  & \{nconst\}                    \\
 & name\_professions & np  & \{nconst, primaryprofession\} \\
 & name\_titles      & nt  & \{nconst, knownfortitle\}     \\
 & title\_akas       & ta  & \{tconst, ordering\}          \\
 & title\_attributes & tat & \{tconst, attribute\}         \\
 & title\_basics     & tb  & \{tconst\}                    \\
 & title\_characters & tc  & \{tconst, nconst, character\} \\
 & title\_directors  & td  & \{tconst, director\}          \\
 & title\_episodes   & te  & \{tconst\}                    \\
 & title\_genres     & tg  & \{tconst, genre\}             \\
 & title\_principals & tp  & \{nconst, tconst, ordering\}  \\
 & title\_types      & ty  & \{tconst, type\}              \\
 & title\_writers    & tw  & \{tconst, writer\}            \\
\hline
\end{tabular}
\end{table*}

\subsection{Queries and DBMS configuration}
\label{sec-setting-optimization}

After all the databases (i.e., baseline, 1NF, 2NF, and 4NF) and their tables had been created, we populated the tables in the 1NF, 2NF, and 4NF databases using the data in the baseline database. Next, we created primary indices for all the tables based on their candidate keys. Next, we considered typical database queries which the DBMS used in the IMDb system would handle. As we had no access to the actual uses cases, nor the whole database in terms of structure and data, these queries are speculative, and based on experiences in using the IMDb web site. As the movie reviews are missing from the dataset, and they are arguably the most write-heavy aspect of the database (i.e., there are significantly more movie reviews added within a time period than, say, movies or people), we chose to limit the scope of our research to read operations. The seven queries formed basis for our tests, and they satisfy data demands such as \textit{``find the best 250 movies based on their rating''} and \textit{``'find the names of the titles a specific person is known for'}. Based on these seven queries, we implemented secondary indices for the tables. As the database structures between the databases differ because of the different normal forms, the queries and consequently the secondary indices differ as well.

For transparency, all the SQL statements used in the creation of the databases can be found in Supplementary material\footnote{https://anonymous.4open.science/r/f4-normalization-74E7} under \texttt{inputs}: table creation statements in the file \texttt{01\_create\_tables.sql} (all databases), table population statements in files starting with \texttt{02}, primary and secondary index creation statements in files starting with \texttt{03} and \texttt{04}, respectively, and queries in files starting with \texttt{05}. The full filenames indicate which database the statements are associated with.

We configured PostgreSQL according to general guidelines\footnote{https://pgtune.leopard.in.ua} based on our hardware. We acknowledge that these guidelines are indeed general, and dependent on many other considerations besides hardware. Prior to the tests proper, we run some preliminary, shorter tests with different configuration parameters -- mainly PostgreSQL's \texttt{shared\_buffers}, \texttt{effective\_cache\_size} and \texttt{work\_mem}, and \texttt{pgbench}'s number of clients -- but observed no major changes towards better results in transactions per second. All the databases are tested with the same configuration and 48 concurrent clients. The full PostgreSQL configuration file can be found in Supplementary material under \texttt{config}. All transactions are run with PostgreSQL's default isolation level (\texttt{READ-COMMITTED}), although this is of little consequence due to the nature of the tests.

\subsection{Determining database size}
\label{sec-setting-size}


We determined database sizes on disk for all the four databases by querying PostgreSQL metadata through system catalogs. This was done with and without table indices, including both primary and secondary indices. The SQL statements for determining database sizes can be found in Supplementary material under \texttt{utils}. We also describe some logical aspects, namely number of tables and number of rows for each database. These numbers were obtained through PostgreSQL metadata and simple \texttt{COUNT(*)} statements.

\subsection{Query complexity measurement}


We report query complexities of the seven queries for each of the respective four databases with query trees. The query complexity consists of the number of selections, tables and table joins in the query. We determined the query complexity by consulting the query execution plans obtained with \texttt{EXPLAIN ANALYZE}. The query execution plans can be found in Supplementary material under \texttt{qeps}.

\subsection{Query performance and energy consumption measurement apparatus}
\label{sec-setting-performance}

We used \texttt{pgbench}\footnote{https://www.postgresql.org/docs/current/pgbench.html} for evaluating throughput. \texttt{pgbench} is a PostgreSQL built-in performance evaluation tool which simulates concurrent transactions, executes SQL statements against a selected database, and logs the results. We used a single machine with no virtualization to test performance. Table~\ref{table-hardware} details relevant components. 

\begin{table}
  \centering
  \caption{Hardware and software used in query performance measurement}
  \label{table-hardware}
  \begin{tabular}{ll}
\toprule
Component  & Description                             \\
\midrule
CPU        & Intel i9-13900K @ 5.80 GHz (24 cores)   \\ 
L1d caches & 896 KB (total of 24 instances)          \\
L1i caches & 1.3 MB (total of 24 instances)          \\
L2 caches  & 32 MB (total of 12 instances)           \\
L3 caches  & 36 MB (one instance)                    \\
Memory     & 64 GB DDR5 4,800 MHz (2 $\times$ 32 GB) \\
Disk       & Seagate 2TB NVMe SDD, PCIe 4.0          \\
OS         & Ubuntu Server 22.04.3 64 bit            \\ 
DBMS       & PostgreSQL 16.1                         \\ 
\bottomrule
\end{tabular}
\end{table}

Additionally, we used \texttt{powerstat} to measure power consumption. \texttt{powerstat} is a software-based measurement tool, and does not differentiate between power consumed \texttt{pgbench}, PostgreSQL or different background processes. Therefore, the numbers reported in the subsequent sections concerning memory, CPU and power consumption account for the whole system. To minimize the effects of different background processes, no additional software was installed, and network connection was disabled during the tests. Each test lasts one hour, and each test was run three times, i.e., we ran 3 $\times$ 4 tests in total. Between each of the 12 test runs, the CPU and OS caches were flushed\footnote{\texttt{\$ sudo echo 3 > /proc/sys/vm/drop\_caches}, as instructed \cite{Raasveldt_2018}}, and the system was hard rebooted. In addition to the tests proper, which measure performance in transaction processing, we also report time and power consumption in database creation for the 1NF, 2NF and 4NF databases, as suggested in a previous study \cite{Raasveldt_2018}. It is worth noting, though, that the IMDb database is not created this way (i.e, in bulk), but rather with titles, crew and reviews being added naturally over time. Since the baseline database is created in a different way than the databases derived from the baseline database, the creation of the baseline database was not measured or compared. The database creation benchmarks were run only once.



\section{Results}
\label{sec-results}

\subsection{Database size}
\label{sec-results-database-size}

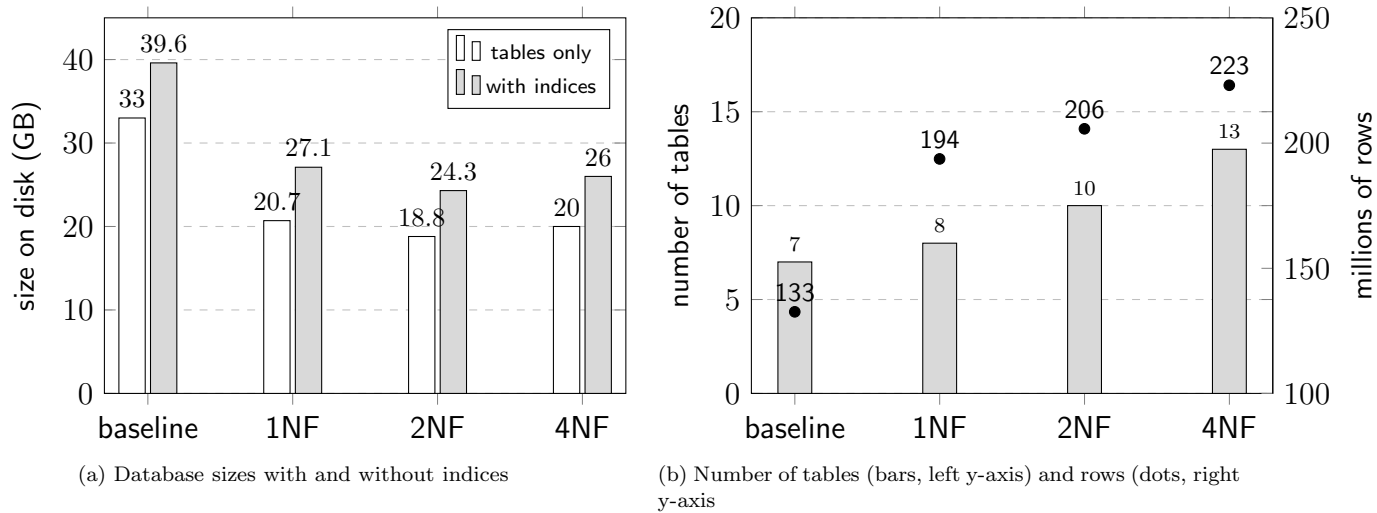
\begin{figure*}
    \centering
\begin{subfigure}[t]{0.44\textwidth}
\centering
        \pgfplotstableread[row sep=\\,col sep=&]{
            interval & tables & indices\\
            baseline & 33.0   & 39.6 \\
            1NF      & 20.7   & 27.1 \\
            2NF      & 18.8   & 24.3 \\
            4NF      & 20.0   & 26.0 \\
        }\mydata
\begin{tikzpicture}
    \begin{axis}[
    ybar,
    bar width=.35cm,
    width=1.1\textwidth,
    height=.85\textwidth,
    symbolic x coords={baseline, 1NF, 2NF, 4NF},
    xtick=data,
    nodes near coords,
    nodes near coords align={vertical},
    ymin=0,ymax=45,
    ylabel=\small size on disk (GB),
    ymajorgrids=true,
    grid style=dashed,
    legend pos=north east,
    every node near coord/.append style={font=\footnotesize},
    ]
    \addplot[style={draw=black,fill=white}] table[x=interval,y=tables]{\mydata};
    \addplot[style={draw=black,fill=light-gray}] table[x=interval,y=indices]{\mydata};
    \legend{\scriptsize tables only,\scriptsize with indices};
    \end{axis}
\end{tikzpicture}
    \caption{Database sizes with and without indices}
    \label{fig-database-size1}
\end{subfigure}
\hspace{0.04\textwidth} 
\begin{subfigure}[t]{0.44\textwidth}  
    \centering 

        \pgfplotstableread[row sep=\\,col sep=&]{
            interval & indices \\
            baseline & 7       \\
            1NF      & 8       \\
            2NF      & 10      \\
            4NF      & 13      \\
        }\mydata
\begin{tikzpicture}
    \begin{axis}[
    ybar,
    bar width=.45cm,
    width=1.1\textwidth,
    height=.85\textwidth,
    symbolic x coords={baseline, 1NF, 2NF, 4NF},
    xtick=data,
    nodes near coords,
    ymin=0,ymax=20,
    ylabel=\small number of tables,
    ymajorgrids=true,
    grid style=dashed,
    every node near coord/.append style={font=\scriptsize},
    ]
    \addplot[style={draw=black,fill=light-gray,font=\footnotesize}] table[,x=interval,y=indices]{\mydata};  
    \end{axis}
    \begin{axis}[
    axis y line*=right,
    axis x line=none,
    width=1.1\textwidth,
    height=.85\textwidth,
    enlarge x limits=0.25,
    axis line style={-},
    ylabel = {\small millions of rows},
    xmin=0, xmax=6,
    scaled y ticks = false,
    ymin=100, ymax=250,
    ]
    \addplot[black,mark=*] table {
    A B
    -0.75 132.624

    1.75 193.688

    4.25 205.716

    6.75 223.111
    }; 
   \node [above] at (axis cs:  -0.75, 132.624) {\footnotesize 133};
   \node [above] at (axis cs:  1.75,  193.688) {\footnotesize 194};
   \node [above] at (axis cs:  4.25,  205.716) {\footnotesize 206};
   \node [above] at (axis cs:  6.75,  223.111) {\footnotesize 223};
    \end{axis}
\end{tikzpicture}

    \caption{Number of tables (bars, left y-axis) and rows (dots, right y-axis}  
    \label{fig-database-size2}
\end{subfigure}
    \caption{Database sizes by several metrics; \textit{baseline} is the original NFNF IMDb database}
    \label{fig-database-size}
\end{figure*}

Fig.~\ref{fig-database-size} describes database sizes by several metrics. Fig.~\ref{fig-database-size1} shows that while the normalization from 1NF to 2NF, indices included, reduces storage space approximately 10\%, the database adhering to 4NF requires 9\% more storage space than the 2NF database. All the normalized database structures require significantly less storage space (reduction ranging from 22\% to 39\%) than the NFNF baseline database, suggesting that the baseline database is subject to redundancy, most likely by design. In contrast, the number of tables as well as table rows in Fig.~\ref{fig-database-size2} grow monotonically with more strict normal forms despite the fact that the amount of required storage space does not. Despite the fact that the baseline database contains approximately 133 million rows and the 4NF database approximately 223 million rows, the average number of rows per table is smaller in the 4NF database. 

\subsection{Query complexity}
\label{sec-results-query-complexity}

\begin{figure*}
    \centering
        \begin{subfigure}[t]{0.20\textwidth}
            \centering
\tikzset{%
    mylabel/.style={font=\normalsize},
    pics/entity/.style n args={3}{code={%
        \node[draw,
        rectangle split,
        rectangle split parts=2,
        text height=1ex,
        text width=9em,
        ] (#1)
        {#2 \nodepart[font=\small]{second}
            \begin{tabular}{>{\raggedright\arraybackslash}p{10em}}
                #3
            \end{tabular}
        };%
    }}
}
\resizebox{0.98\linewidth}{!}{
\begin{tikzpicture}

\draw[xstep=1,ystep=1,white,thin] (-0.5,-0.5) grid (1.5,2.5);
\draw (current bounding box.north east) -- (current bounding box.north west) -- (current bounding box.south west) -- (current bounding box.south east) -- cycle;

\draw (0,0) circle [radius=0.23] node(tb)  {tb};
\draw (1,0) circle [radius=0.23] node(tr)  {tr};

\draw (0.5, 1) node(bt1) {\LARGE$\bowtie$};
\draw (0.5, 2) node(sel) {\LARGE$\pi$};

\draw[->] (tb)  to (bt1) node[yshift=-1.4em, xshift=-1.5em]{$\sigma$};
\draw[->] (tr)  to (bt1) node[yshift=-1.4em, xshift=1.5em]{$\sigma$};
\draw[->] (bt1) to (sel);

\end{tikzpicture}
}
            \vspace{-1\baselineskip}
            \caption[]%
            {{\small Query \#1, bl}}
            \label{fig-q1-bl}
        \end{subfigure}
        \begin{subfigure}[t]{0.20\textwidth}  
            \centering 
\tikzset{%
    mylabel/.style={font=\normalsize},
    pics/entity/.style n args={3}{code={%
        \node[draw,
        rectangle split,
        rectangle split parts=2,
        text height=1ex,
        text width=9em,
        ] (#1)
        {#2 \nodepart[font=\small]{second}
            \begin{tabular}{>{\raggedright\arraybackslash}p{10em}}
                #3
            \end{tabular}
        };%
    }}
}
\resizebox{0.98\linewidth}{!}{
\begin{tikzpicture}

\draw[xstep=1,ystep=1,white,thin] (-0.5,-0.5) grid (1.5,2.5);
\draw (current bounding box.north east) -- (current bounding box.north west) -- (current bounding box.south west) -- (current bounding box.south east) -- cycle;

\draw (0.5,1) circle [radius=0.23] node(tb)  {tb};

\draw (0.5, 2) node(sel) {\LARGE$\pi$};

\draw[->] (tb)  to (sel) node[yshift=-1.4em, xshift=-1em]{$\sigma$};

\end{tikzpicture}
}
            \vspace{-1\baselineskip}
            \caption[]%
            {{\small Query \#1, 1NF}}    
            \label{fig-q1-1nf}
        \end{subfigure}
        \begin{subfigure}[t]{0.20\textwidth}  
            \centering 
\tikzset{%
    mylabel/.style={font=\normalsize},
    pics/entity/.style n args={3}{code={%
        \node[draw,
        rectangle split,
        rectangle split parts=2,
        text height=1ex,
        text width=9em,
        ] (#1)
        {#2 \nodepart[font=\small]{second}
            \begin{tabular}{>{\raggedright\arraybackslash}p{10em}}
                #3
            \end{tabular}
        };%
    }}
}
\resizebox{0.98\linewidth}{!}{
\begin{tikzpicture}

\draw[xstep=1,ystep=1,white,thin] (-0.5,-0.5) grid (1.5,2.5);
\draw (current bounding box.north east) -- (current bounding box.north west) -- (current bounding box.south west) -- (current bounding box.south east) -- cycle;

\draw (0,0) circle [radius=0.23] node(tb)  {tb};
\draw (1,0) circle [radius=0.23] node(ta)  {ta};

\draw (0.5, 1) node(bt1) {\LARGE$\bowtie$};
\draw (0.5, 2) node(sel) {\LARGE$\pi$};

\draw[->] (tb)  to (bt1) node[yshift=-1.4em, xshift=-1.5em]{$\sigma$};
\draw[->] (ta)  to (bt1) node[yshift=-1.4em, xshift=1.5em]{$\sigma$};
\draw[->] (bt1) to (sel);

\end{tikzpicture}
}
            \vspace{-1\baselineskip}
            \caption[]%
            {{\small Query \#1, 2NF}}    
            \label{fig-q1-2nf}
        \end{subfigure}
        \begin{subfigure}[t]{0.20\textwidth}  
            \centering 
\tikzset{%
    mylabel/.style={font=\normalsize},
    pics/entity/.style n args={3}{code={%
        \node[draw,
        rectangle split,
        rectangle split parts=2,
        text height=1ex,
        text width=9em,
        ] (#1)
        {#2 \nodepart[font=\small]{second}
            \begin{tabular}{>{\raggedright\arraybackslash}p{10em}}
                #3
            \end{tabular}
        };%
    }}
}
\resizebox{0.98\linewidth}{!}{
\begin{tikzpicture}

\draw[xstep=1,ystep=1,white,thin] (-0.5,-0.5) grid (1.5,2.5);
\draw (current bounding box.north east) -- (current bounding box.north west) -- (current bounding box.south west) -- (current bounding box.south east) -- cycle;

\draw (0,0) circle [radius=0.23] node(tb)  {tb};
\draw (1,0) circle [radius=0.23] node(ta)  {ta};

\draw (0.5, 1) node(bt1) {\LARGE$\bowtie$};
\draw (0.5, 2) node(sel) {\LARGE$\pi$};

\draw[->] (tb)  to (bt1) node[yshift=-1.4em, xshift=-1.5em]{$\sigma$};
\draw[->] (ta)  to (bt1) node[yshift=-1.4em, xshift=1.5em]{$\sigma$};
\draw[->] (bt1) to (sel);

\end{tikzpicture}
}
            \vspace{-1\baselineskip}
            \caption[]%
            {{\small Query \#1, 4NF}}    
            \label{fig-q1-4nf}
        \end{subfigure}
        \vskip\baselineskip 
        \begin{subfigure}[t]{0.20\textwidth}   
            \centering 
\tikzset{%
    mylabel/.style={font=\normalsize},
    pics/entity/.style n args={3}{code={%
        \node[draw,
        rectangle split,
        rectangle split parts=2,
        text height=1ex,
        text width=9em,
        ] (#1)
        {#2 \nodepart[font=\small]{second}
            \begin{tabular}{>{\raggedright\arraybackslash}p{10em}}
                #3
            \end{tabular}
        };%
    }}
}
\resizebox{0.98\linewidth}{!}{
\begin{tikzpicture}

\draw[xstep=1,ystep=1,white,thin] (-0.5,-0.5) grid (1.5,2.5);
\draw (current bounding box.north east) -- (current bounding box.north west) -- (current bounding box.south west) -- (current bounding box.south east) -- cycle;

\draw (0,0) circle [radius=0.23] node(tb)  {tb};
\draw (1,0) circle [radius=0.23] node(nb)  {nb};

\draw (0.5, 1) node(bt1) {\LARGE$\bowtie$};
\draw (0.5, 2) node(sel) {\LARGE$\pi$};

\draw[->] (tb)  to (bt1);
\draw[->] (nb)  to (bt1) node[yshift=-1.4em, xshift=1.5em]{$\sigma$};
\draw[->] (bt1) to (sel);

\end{tikzpicture}
}
            \vspace{-1\baselineskip}
            \caption[]%
            {{\small Query \#2, bl}}    
            \label{fig-q2-bl}
        \end{subfigure}
        \begin{subfigure}[t]{0.20\textwidth}   
            \centering 
\tikzset{%
    mylabel/.style={font=\normalsize},
    pics/entity/.style n args={3}{code={%
        \node[draw,
        rectangle split,
        rectangle split parts=2,
        text height=1ex,
        text width=9em,
        ] (#1)
        {#2 \nodepart[font=\small]{second}
            \begin{tabular}{>{\raggedright\arraybackslash}p{10em}}
                #3
            \end{tabular}
        };%
    }}
}
\resizebox{0.98\linewidth}{!}{
\begin{tikzpicture}

\draw[xstep=1,ystep=1,white,thin] (-0.5,-0.5) grid (1.5,2.5);
\draw (current bounding box.north east) -- (current bounding box.north west) -- (current bounding box.south west) -- (current bounding box.south east) -- cycle;

\draw (0,0) circle [radius=0.23] node(tb)  {tb};
\draw (1,0) circle [radius=0.23] node(npt)  {\scriptsize npt};

\draw (0.5, 1) node(bt1) {\LARGE$\bowtie$};
\draw (0.5, 2) node(sel) {\LARGE$\pi$};

\draw[->] (tb)  to (bt1) node[yshift=-1.4em, xshift=-1.5em]{$\sigma$};
\draw[->] (npt)  to (bt1) node[yshift=-1.4em, xshift=1.5em]{$\sigma$};
\draw[->] (bt1) to (sel);

\end{tikzpicture}
}
            \vspace{-1\baselineskip}
            \caption[]%
            {{\small Query \#2, 1NF}}    
            \label{fig-q2-1nf}
        \end{subfigure}
        \begin{subfigure}[t]{0.20\textwidth}   
            \centering 
\tikzset{%
    mylabel/.style={font=\normalsize},
    pics/entity/.style n args={3}{code={%
        \node[draw,
        rectangle split,
        rectangle split parts=2,
        text height=1ex,
        text width=9em,
        ] (#1)
        {#2 \nodepart[font=\small]{second}
            \begin{tabular}{>{\raggedright\arraybackslash}p{10em}}
                #3
            \end{tabular}
        };%
    }}
}
\resizebox{0.98\linewidth}{!}{
\begin{tikzpicture}

\draw[xstep=1,ystep=1,white,thin] (-0.5,-0.5) grid (1.5,2.5);
\draw (current bounding box.north east) -- (current bounding box.north west) -- (current bounding box.south west) -- (current bounding box.south east) -- cycle;

\draw (0,0) circle [radius=0.23] node(ta)  {ta};
\draw (1,0) circle [radius=0.23] node(npt)  {\scriptsize npt};

\draw (0.5, 1) node(bt1) {\LARGE$\bowtie$};
\draw (0.5, 2) node(sel) {\LARGE$\pi$};

\draw[->] (ta)  to (bt1) node[yshift=-1.4em, xshift=-1.5em]{$\sigma$};
\draw[->] (npt)  to (bt1) node[yshift=-1.4em, xshift=1.5em]{$\sigma$};
\draw[->] (bt1) to (sel);

\end{tikzpicture}
}
            \vspace{-1\baselineskip}
            \caption[]%
            {{\small Query \#2, 2NF}}    
            \label{fig-q2-2nf}
        \end{subfigure}
        \begin{subfigure}[t]{0.20\textwidth}   
            \centering 
\tikzset{%
    mylabel/.style={font=\normalsize},
    pics/entity/.style n args={3}{code={%
        \node[draw,
        rectangle split,
        rectangle split parts=2,
        text height=1ex,
        text width=9em,
        ] (#1)
        {#2 \nodepart[font=\small]{second}
            \begin{tabular}{>{\raggedright\arraybackslash}p{10em}}
                #3
            \end{tabular}
        };%
    }}
}
\resizebox{0.98\linewidth}{!}{
\begin{tikzpicture}

\draw[xstep=1,ystep=1,white,thin] (-0.5,-0.5) grid (1.5,2.5);
\draw (current bounding box.north east) -- (current bounding box.north west) -- (current bounding box.south west) -- (current bounding box.south east) -- cycle;

\draw (0,0) circle [radius=0.23] node(ta) {ta};
\draw (1,0) circle [radius=0.23] node(nt) {nt};

\draw (0.5, 1) node(bt1) {\LARGE$\bowtie$};
\draw (0.5, 2) node(sel) {\LARGE$\pi$};

\draw[->] (ta)  to (bt1) node[yshift=-1.4em, xshift=-1.5em]{$\sigma$};
\draw[->] (nt)  to (bt1) node[yshift=-1.4em, xshift=1.5em]{$\sigma$};
\draw[->] (bt1) to (sel);

\end{tikzpicture}
}
            \vspace{-1\baselineskip}
            \caption[]%
            {{\small Query \#2, 4NF}}    
            \label{fig-q2-4nf}
        \end{subfigure}
        \vskip\baselineskip 
        \begin{subfigure}[t]{0.20\textwidth}   
            \centering 
\tikzset{%
    mylabel/.style={font=\normalsize},
    pics/entity/.style n args={3}{code={%
        \node[draw,
        rectangle split,
        rectangle split parts=2,
        text height=1ex,
        text width=9em,
        ] (#1)
        {#2 \nodepart[font=\small]{second}
            \begin{tabular}{>{\raggedright\arraybackslash}p{10em}}
                #3
            \end{tabular}
        };%
    }}
}
\resizebox{0.98\linewidth}{!}{
\begin{tikzpicture}

\draw[xstep=1,ystep=1,white,thin] (-0.75,-0.5) grid (1.75,3.5);
\draw (current bounding box.north east) -- (current bounding box.north west) -- (current bounding box.south west) -- (current bounding box.south east) -- cycle;

\draw (0,1) circle [radius=0.23] node(tb)  {tb};
\draw (1,1) circle [radius=0.23] node(te)  {te};

\draw (0.5, 2) node(bt2) {\LARGE$\bowtie$};
\draw (0.5, 3) node(sel) {\LARGE$\pi$};

\draw[->] (tb)  to (bt2) node[yshift=-1.4em, xshift=-1.5em]{$\sigma$};
\draw[->] (te) to  (bt2);
\draw[->] (bt2) to (sel);

\end{tikzpicture}
}
            \vspace{-1\baselineskip}
            \caption[]%
            {{\small Query \#3, bl}}    
            \label{fig-q3-bl}
        \end{subfigure}
        \begin{subfigure}[t]{0.20\textwidth}   
            \centering 
\tikzset{%
    mylabel/.style={font=\normalsize},
    pics/entity/.style n args={3}{code={%
        \node[draw,
        rectangle split,
        rectangle split parts=2,
        text height=1ex,
        text width=9em,
        ] (#1)
        {#2 \nodepart[font=\small]{second}
            \begin{tabular}{>{\raggedright\arraybackslash}p{10em}}
                #3
            \end{tabular}
        };%
    }}
}
\resizebox{0.98\linewidth}{!}{
\begin{tikzpicture}

\draw[xstep=1,ystep=1,white,thin] (-0.75,-0.5) grid (1.75,3.5);
\draw (current bounding box.north east) -- (current bounding box.north west) -- (current bounding box.south west) -- (current bounding box.south east) -- cycle;

\draw (0,1) circle [radius=0.23] node(tb)  {tb};
\draw (1,1) circle [radius=0.23] node(te)  {te};

\draw (0.5, 2) node(bt2) {\LARGE$\bowtie$};
\draw (0.5, 3) node(sel) {\LARGE$\pi$};

\draw[->] (tb)  to (bt2) node[yshift=-1.4em, xshift=-1.5em]{$\sigma$};
\draw[->] (te) to  (bt2);
\draw[->] (bt2) to (sel);

\end{tikzpicture}
}
            \vspace{-1\baselineskip}
            \caption[]%
            {{\small Query \#3, 1NF}}    
            \label{fig-q3-1nf}
        \end{subfigure}
        \begin{subfigure}[t]{0.20\textwidth}   
            \centering 
\tikzset{%
    mylabel/.style={font=\normalsize},
    pics/entity/.style n args={3}{code={%
        \node[draw,
        rectangle split,
        rectangle split parts=2,
        text height=1ex,
        text width=9em,
        ] (#1)
        {#2 \nodepart[font=\small]{second}
            \begin{tabular}{>{\raggedright\arraybackslash}p{10em}}
                #3
            \end{tabular}
        };%
    }}
}
\resizebox{0.98\linewidth}{!}{
\begin{tikzpicture}

\draw[xstep=1,ystep=1,white,thin] (-0.5,-0.5) grid (2,3.5);
\draw (current bounding box.north east) -- (current bounding box.north west) -- (current bounding box.south west) -- (current bounding box.south east) -- cycle;

\draw (0,0) circle [radius=0.23] node(tb)  {tb};
\draw (1,0) circle [radius=0.23] node(ta)  {ta};
\draw (1.5,1) circle [radius=0.23] node(te)  {te};

\draw (0.5, 1) node(bt1) {\LARGE$\bowtie$};
\draw (0.5, 2) node(bt2) {\LARGE$\bowtie$};
\draw (0.5, 3) node(sel) {\LARGE$\pi$};

\draw[->] (tb)  to (bt1) node[yshift=-1.4em, xshift=-1.5em]{$\sigma$};
\draw[->] (ta)  to (bt1);
\draw[->] (te)  to (bt2);
\draw[->] (bt1) to (bt2);
\draw[->] (bt2) to (sel);

\end{tikzpicture}
}
            \vspace{-1\baselineskip}
            \caption[]%
            {{\small Query \#3, 2NF}}    
            \label{fig-q3-2nf}
        \end{subfigure}
        \begin{subfigure}[t]{0.20\textwidth}   
            \centering 
\tikzset{%
    mylabel/.style={font=\normalsize},
    pics/entity/.style n args={3}{code={%
        \node[draw,
        rectangle split,
        rectangle split parts=2,
        text height=1ex,
        text width=9em,
        ] (#1)
        {#2 \nodepart[font=\small]{second}
            \begin{tabular}{>{\raggedright\arraybackslash}p{10em}}
                #3
            \end{tabular}
        };%
    }}
}
\resizebox{0.98\linewidth}{!}{
\begin{tikzpicture}

\draw[xstep=1,ystep=1,white,thin] (-0.5,-0.5) grid (2,3.5);
\draw (current bounding box.north east) -- (current bounding box.north west) -- (current bounding box.south west) -- (current bounding box.south east) -- cycle;

\draw (0,0) circle [radius=0.23] node(tb)  {tb};
\draw (1,0) circle [radius=0.23] node(ta)  {ta};
\draw (1.5,1) circle [radius=0.23] node(te)  {te};

\draw (0.5, 1) node(bt1) {\LARGE$\bowtie$};
\draw (0.5, 2) node(bt2) {\LARGE$\bowtie$};
\draw (0.5, 3) node(sel) {\LARGE$\pi$};

\draw[->] (tb)  to (bt1) node[yshift=-1.4em, xshift=-1.5em]{$\sigma$};
\draw[->] (ta)  to (bt1);
\draw[->] (te)  to (bt2);
\draw[->] (bt1) to (bt2);
\draw[->] (bt2) to (sel);

\end{tikzpicture}
}
            \vspace{-1\baselineskip}
            \caption[]%
            {{\small Query \#3, 4NF}}
            \label{fig-q3-4nf}
        \end{subfigure}
        \vskip\baselineskip 
        \begin{subfigure}[t]{0.20\textwidth}   
            \centering 
\tikzset{%
    mylabel/.style={font=\normalsize},
    pics/entity/.style n args={3}{code={%
        \node[draw,
        rectangle split,
        rectangle split parts=2,
        text height=1ex,
        text width=9em,
        ] (#1)
        {#2 \nodepart[font=\small]{second}
            \begin{tabular}{>{\raggedright\arraybackslash}p{10em}}
                #3
            \end{tabular}
        };%
    }}
}
\resizebox{0.98\linewidth}{!}{
\begin{tikzpicture}

\draw[xstep=1,ystep=1,white,thin] (-0.75,-0.5) grid (1.75,3.5);
\draw (current bounding box.north east) -- (current bounding box.north west) -- (current bounding box.south west) -- (current bounding box.south east) -- cycle;

\draw (0.5,2) circle [radius=0.23] node(nb)  {nb};

\draw (0.5, 3) node(sel) {\LARGE$\pi$};

\draw[->] (nb)  to (sel) node[yshift=-1.4em, xshift=-1.1em]{$\sigma$};

\end{tikzpicture}
}
            \vspace{-1\baselineskip}
            \caption[]%
            {{\small Query \#4, bl}}    
            \label{fig-q4-bl}
        \end{subfigure}
        \begin{subfigure}[t]{0.20\textwidth}   
            \centering 
\tikzset{%
    mylabel/.style={font=\normalsize},
    pics/entity/.style n args={3}{code={%
        \node[draw,
        rectangle split,
        rectangle split parts=2,
        text height=1ex,
        text width=9em,
        ] (#1)
        {#2 \nodepart[font=\small]{second}
            \begin{tabular}{>{\raggedright\arraybackslash}p{10em}}
                #3
            \end{tabular}
        };%
    }}
}
\resizebox{0.98\linewidth}{!}{
\begin{tikzpicture}

\draw[xstep=1,ystep=1,white,thin] (-0.75,-0.5) grid (1.75,3.5);
\draw (current bounding box.north east) -- (current bounding box.north west) -- (current bounding box.south west) -- (current bounding box.south east) -- cycle;

\draw (0,1) circle [radius=0.23] node(nb)  {nb};
\draw (1,1) circle [radius=0.23] node(npt)  {\scriptsize npt};

\draw (0.5, 2) node(bt1) {\LARGE$\leftouterjoin$};
\draw (0.5, 3) node(sel) {\LARGE$\pi$};

\draw[->] (nb)  to (bt1) node[yshift=-1.4em, xshift=-1.5em]{$\sigma$};
\draw[->] (npt) to (bt1);
\draw[->] (bt1) to (sel);

\end{tikzpicture}
}
            \vspace{-1\baselineskip}
            \caption[]%
            {{\small Query \#4, 1NF}}    
            \label{fig-q4-1nf}
        \end{subfigure}
        \begin{subfigure}[t]{0.20\textwidth}   
            \centering 
\tikzset{%
    mylabel/.style={font=\normalsize},
    pics/entity/.style n args={3}{code={%
        \node[draw,
        rectangle split,
        rectangle split parts=2,
        text height=1ex,
        text width=9em,
        ] (#1)
        {#2 \nodepart[font=\small]{second}
            \begin{tabular}{>{\raggedright\arraybackslash}p{10em}}
                #3
            \end{tabular}
        };%
    }}
}
\resizebox{0.98\linewidth}{!}{
\begin{tikzpicture}

\draw[xstep=1,ystep=1,white,thin] (-0.75,-0.5) grid (1.75,3.5);
\draw (current bounding box.north east) -- (current bounding box.north west) -- (current bounding box.south west) -- (current bounding box.south east) -- cycle;

\draw (0,1) circle [radius=0.23] node(nb)  {nb};
\draw (1,1) circle [radius=0.23] node(npt)  {\scriptsize npt};

\draw (0.5, 2) node(bt1) {\LARGE$\leftouterjoin$};
\draw (0.5, 3) node(sel) {\LARGE$\pi$};

\draw[->] (nb)  to (bt1) node[yshift=-1.4em, xshift=-1.5em]{$\sigma$};
\draw[->] (npt) to (bt1);
\draw[->] (bt1) to (sel);

\end{tikzpicture}
}
            \vspace{-1\baselineskip}
            \caption[]%
            {{\small Query \#4, 2NF}}    
            \label{fig-q4-2nf}
        \end{subfigure}
        \begin{subfigure}[t]{0.20\textwidth}   
            \centering 
\tikzset{%
    mylabel/.style={font=\normalsize},
    pics/entity/.style n args={3}{code={%
        \node[draw,
        rectangle split,
        rectangle split parts=2,
        text height=1ex,
        text width=9em,
        ] (#1)
        {#2 \nodepart[font=\small]{second}
            \begin{tabular}{>{\raggedright\arraybackslash}p{10em}}
                #3
            \end{tabular}
        };%
    }}
}
\resizebox{0.98\linewidth}{!}{
\begin{tikzpicture}

\draw[xstep=1,ystep=1,white,thin] (-0.5,-0.5) grid (2,3.5);
\draw (current bounding box.north east) -- (current bounding box.north west) -- (current bounding box.south west) -- (current bounding box.south east) -- cycle;

\draw (0,0) circle [radius=0.23] node(nb)  {nb};
\draw (1,0) circle [radius=0.23] node(npt)  {\scriptsize npt};
\draw (1.5,1) circle [radius=0.23] node(nt) {nt};

\draw (0.5, 1) node(bt1) {\LARGE$\leftouterjoin$};
\draw (0.5, 2) node(bt2) {\LARGE$\leftouterjoin$};
\draw (0.5, 3) node(sel) {\LARGE$\pi$};

\draw[->] (nb)  to (bt1) node[yshift=-1.4em, xshift=-1.5em]{$\sigma$};
\draw[->] (npt) to (bt1);
\draw[->] (nt)  to (bt2);
\draw[->] (bt1) to (bt2);
\draw[->] (bt2) to (sel);

\end{tikzpicture}
}
            \vspace{-1\baselineskip}
            \caption[]%
            {{\small Query \#4, 4NF}}    
            \label{fig-q4-4nf}
        \end{subfigure}
    \caption{Query trees for queries 1..4; nodes represent tables, bowties joins, $\sigma$s selection and $\pi$s projection} 
    \label{fig-query-trees-1}
\end{figure*}

\begin{figure*}
    \centering
        \begin{subfigure}[t]{0.2\textwidth}
            \centering
\tikzset{%
    mylabel/.style={font=\normalsize},
    pics/entity/.style n args={3}{code={%
        \node[draw,
        rectangle split,
        rectangle split parts=2,
        text height=1ex,
        text width=9em,
        ] (#1)
        {#2 \nodepart[font=\small]{second}
            \begin{tabular}{>{\raggedright\arraybackslash}p{10em}}
                #3
            \end{tabular}
        };%
    }}
}
\resizebox{0.98\linewidth}{!}{
\begin{tikzpicture}

\draw[xstep=1,ystep=1,white,thin] (-0.75,-0.5) grid (1.75,3.5);
\draw (current bounding box.north east) -- (current bounding box.north west) -- (current bounding box.south west) -- (current bounding box.south east) -- cycle;

\draw (0.5,2) circle [radius=0.23] node(ta)  {ta};

\draw (0.5, 3) node(sel) {\LARGE$\pi$};

\draw[->] (ta)  to (sel) node[yshift=-1.4em, xshift=-1.1em]{$\sigma$};

\end{tikzpicture}
}
            \vspace{-1\baselineskip}
            \caption[]%
            {{\small Query \#5, baseline}}
            \label{fig-q5-bl}
        \end{subfigure}
        \begin{subfigure}[t]{0.2\textwidth}  
            \centering 
\tikzset{%
    mylabel/.style={font=\normalsize},
    pics/entity/.style n args={3}{code={%
        \node[draw,
        rectangle split,
        rectangle split parts=2,
        text height=1ex,
        text width=9em,
        ] (#1)
        {#2 \nodepart[font=\small]{second}
            \begin{tabular}{>{\raggedright\arraybackslash}p{10em}}
                #3
            \end{tabular}
        };%
    }}
}
\resizebox{0.98\linewidth}{!}{
\begin{tikzpicture}

\draw[xstep=1,ystep=1,white,thin] (-0.75,-0.5) grid (1.75,3.5);
\draw (current bounding box.north east) -- (current bounding box.north west) -- (current bounding box.south west) -- (current bounding box.south east) -- cycle;

\draw (0,1) circle [radius=0.23] node(tb)  {tb};
\draw (1,1) circle [radius=0.23] node(tya)  {\scriptsize tya};

\draw (0.5, 2) node(bt1) {\LARGE$\bowtie$};
\draw (0.5, 3) node(sel) {\LARGE$\pi$};

\draw[->] (tb)  to (bt1) node[yshift=-1.4em, xshift=-1.5em]{$\sigma$};
\draw[->] (tya) to (bt1);
\draw[->] (bt1) to (sel);

\end{tikzpicture}
}
            \vspace{-1\baselineskip}
            \caption[]%
            {{\small Query \#5, 1NF}}    
            \label{fig-q5-1nf}
        \end{subfigure}
        \begin{subfigure}[t]{0.2\textwidth}  
            \centering 
\tikzset{%
    mylabel/.style={font=\normalsize},
    pics/entity/.style n args={3}{code={%
        \node[draw,
        rectangle split,
        rectangle split parts=2,
        text height=1ex,
        text width=9em,
        ] (#1)
        {#2 \nodepart[font=\small]{second}
            \begin{tabular}{>{\raggedright\arraybackslash}p{10em}}
                #3
            \end{tabular}
        };%
    }}
}
\resizebox{0.98\linewidth}{!}{
\begin{tikzpicture}

\draw[xstep=1,ystep=1,white,thin] (-0.75,-0.5) grid (1.75,3.5);
\draw (current bounding box.north east) -- (current bounding box.north west) -- (current bounding box.south west) -- (current bounding box.south east) -- cycle;

\draw (0,1) circle [radius=0.23] node(ta)  {ta};
\draw (1,1) circle [radius=0.23] node(tya)  {\scriptsize tya};

\draw (0.5, 2) node(bt1) {\LARGE$\bowtie$};
\draw (0.5, 3) node(sel) {\LARGE$\pi$};

\draw[->] (ta)  to (bt1) node[yshift=-1.4em, xshift=-1.5em]{$\sigma$};
\draw[->] (tya) to (bt1);
\draw[->] (bt1) to (sel);

\end{tikzpicture}
}
            \vspace{-1\baselineskip}
            \caption[]%
            {{\small Query \#5, 2NF}}    
            \label{fig-q5-2nf}
        \end{subfigure}
        \begin{subfigure}[t]{0.2\textwidth}  
            \centering 
\tikzset{%
    mylabel/.style={font=\normalsize},
    pics/entity/.style n args={3}{code={%
        \node[draw,
        rectangle split,
        rectangle split parts=2,
        text height=1ex,
        text width=9em,
        ] (#1)
        {#2 \nodepart[font=\small]{second}
            \begin{tabular}{>{\raggedright\arraybackslash}p{10em}}
                #3
            \end{tabular}
        };%
    }}
}
\resizebox{0.98\linewidth}{!}{
\begin{tikzpicture}

\draw[xstep=1,ystep=1,white,thin] (-0.5,-0.5) grid (2,3.5);
\draw (current bounding box.north east) -- (current bounding box.north west) -- (current bounding box.south west) -- (current bounding box.south east) -- cycle;

\draw (0,0) circle [radius=0.23] node(ty)  {ty};
\draw (1,0) circle [radius=0.23] node(tat)  {\scriptsize tat};
\draw (1.5,1) circle [radius=0.23] node(ta) {ta};

\draw (0.5, 1) node(bt1) {\LARGE$\bowtie$};
\draw (0.5, 2) node(bt2) {\LARGE$\bowtie$};
\draw (0.5, 3) node(sel) {\LARGE$\pi$};

\draw[->] (ty)  to (bt1);
\draw[->] (tat) to (bt1);
\draw[->] (ta)  to (bt2) node[yshift=-1.4em, xshift=2.5em]{$\sigma$};
\draw[->] (bt1) to (bt2);
\draw[->] (bt2) to (sel);

\end{tikzpicture}
}
            \vspace{-1\baselineskip}
            \caption[]%
            {{\small Query \#5, 4NF}}    
            \label{fig-q5-4nf}
        \end{subfigure}
        \\
        \begin{subfigure}[t]{0.2\textwidth}
            \centering
\tikzset{%
    mylabel/.style={font=\normalsize},
    pics/entity/.style n args={3}{code={%
        \node[draw,
        rectangle split,
        rectangle split parts=2,
        text height=1ex,
        text width=9em,
        ] (#1)
        {#2 \nodepart[font=\small]{second}
            \begin{tabular}{>{\raggedright\arraybackslash}p{10em}}
                #3
            \end{tabular}
        };%
    }}
}
\resizebox{0.98\linewidth}{!}{
\begin{tikzpicture}

\draw[xstep=1,ystep=1,white,thin] (-0.5,-0.5) grid (2,3.5);
\draw (current bounding box.north east) -- (current bounding box.north west) -- (current bounding box.south west) -- (current bounding box.south east) -- cycle;

\draw (0,0) circle [radius=0.23] node(tp)   {tp};
\draw (1,0) circle [radius=0.23] node(nb)   {nb};
\draw (1.5,1) circle [radius=0.23] node(tb) {tb};

\draw (0.5, 1) node(bt1) {\LARGE$\bowtie$};
\draw (0.5, 2) node(bt2) {\LARGE$\bowtie$};
\draw (0.5, 3) node(sel) {\LARGE$\pi$};

\draw[->] (tp)  to (bt1) node[yshift=-1.4em, xshift=1.5em]{$\sigma$};
\draw[->] (nb)  to (bt1);
\draw[->] (tb)  to (bt2);
\draw[->] (bt1) to (bt2);
\draw[->] (bt2) to (sel);

\end{tikzpicture}
}
            \vspace{-1\baselineskip}
            \caption[]%
            {{\small Query \#6, baseline}}
            \label{fig-q6-bl}
        \end{subfigure}
        \begin{subfigure}[t]{0.2\textwidth}  
            \centering 
\tikzset{%
    mylabel/.style={font=\normalsize},
    pics/entity/.style n args={3}{code={%
        \node[draw,
        rectangle split,
        rectangle split parts=2,
        text height=1ex,
        text width=9em,
        ] (#1)
        {#2 \nodepart[font=\small]{second}
            \begin{tabular}{>{\raggedright\arraybackslash}p{10em}}
                #3
            \end{tabular}
        };%
    }}
}
\resizebox{0.98\linewidth}{!}{
\begin{tikzpicture}

\draw[xstep=1,ystep=1,white,thin] (-0.75,-0.5) grid (1.75,3.5);
\draw (current bounding box.north east) -- (current bounding box.north west) -- (current bounding box.south west) -- (current bounding box.south east) -- cycle;

\draw (0,1) circle [radius=0.23] node(tb)   {tb};
\draw (1,1) circle [radius=0.23] node(nb)   {nb};

\draw (0.5, 2) node(bt1) {\LARGE$\bowtie$};
\draw (0.5, 3) node(sel) {\LARGE$\pi$};

\draw[->] (tb)  to (bt1) node[yshift=-1.4em, xshift=1.5em]{$\sigma$};
\draw[->] (nb)  to (bt1);
\draw[->] (bt1) to (sel);

\end{tikzpicture}
}
            \vspace{-1\baselineskip}
            \caption[]%
            {{\small Query \#6, 1NF}}    
            \label{fig-q6-1nf}
        \end{subfigure}
        \begin{subfigure}[t]{0.2\textwidth}  
            \centering 
\tikzset{%
    mylabel/.style={font=\normalsize},
    pics/entity/.style n args={3}{code={%
        \node[draw,
        rectangle split,
        rectangle split parts=2,
        text height=1ex,
        text width=9em,
        ] (#1)
        {#2 \nodepart[font=\small]{second}
            \begin{tabular}{>{\raggedright\arraybackslash}p{10em}}
                #3
            \end{tabular}
        };%
    }}
}
\resizebox{0.98\linewidth}{!}{
\begin{tikzpicture}

\draw[xstep=1,ystep=1,white,thin] (-0.5,-0.5) grid (2,3.5);
\draw (current bounding box.north east) -- (current bounding box.north west) -- (current bounding box.south west) -- (current bounding box.south east) -- cycle;

\draw (0,0) circle [radius=0.23] node(tp)   {tp};
\draw (1,0) circle [radius=0.23] node(tb)   {tb};
\draw (1.5,1) circle [radius=0.23] node(nb) {nb};

\draw (0.5, 1) node(bt1) {\LARGE$\bowtie$};
\draw (0.5, 2) node(bt2) {\LARGE$\bowtie$};
\draw (0.5, 3) node(sel) {\LARGE$\pi$};

\draw[->] (tb)  to (bt1) node[yshift=-1.4em, xshift=1.5em]{$\sigma$};
\draw[->] (tp)  to (bt1);
\draw[->] (nb)  to (bt2);
\draw[->] (bt1) to (bt2);
\draw[->] (bt2) to (sel);

\end{tikzpicture}
}
            \vspace{-1\baselineskip}
            \caption[]%
            {{\small Query \#6, 2NF}}    
            \label{fig-q6-2nf}
        \end{subfigure}
        \begin{subfigure}[t]{0.2\textwidth}  
            \centering 
\tikzset{%
    mylabel/.style={font=\normalsize},
    pics/entity/.style n args={3}{code={%
        \node[draw,
        rectangle split,
        rectangle split parts=2,
        text height=1ex,
        text width=9em,
        ] (#1)
        {#2 \nodepart[font=\small]{second}
            \begin{tabular}{>{\raggedright\arraybackslash}p{10em}}
                #3
            \end{tabular}
        };%
    }}
}
\resizebox{0.98\linewidth}{!}{
\begin{tikzpicture}

\draw[xstep=1,ystep=1,white,thin] (-0.5,-0.5) grid (2,3.5);
\draw (current bounding box.north east) -- (current bounding box.north west) -- (current bounding box.south west) -- (current bounding box.south east) -- cycle;

\draw (0,0) circle [radius=0.23] node(tp)   {tp};
\draw (1,0) circle [radius=0.23] node(tb)   {tb};
\draw (1.5,1) circle [radius=0.23] node(nb) {nb};

\draw (0.5, 1) node(bt1) {\LARGE$\bowtie$};
\draw (0.5, 2) node(bt2) {\LARGE$\bowtie$};
\draw (0.5, 3) node(sel) {\LARGE$\pi$};

\draw[->] (tb)  to (bt1) node[yshift=-1.4em, xshift=1.5em]{$\sigma$};
\draw[->] (tp)  to (bt1);
\draw[->] (nb)  to (bt2);
\draw[->] (bt1) to (bt2);
\draw[->] (bt2) to (sel);

\end{tikzpicture}
}
            \vspace{-1\baselineskip}
            \caption[]%
            {{\small Query \#6, 4NF}}    
            \label{fig-q6-4nf}
        \end{subfigure}
        \vskip\baselineskip 
        \begin{subfigure}[t]{0.24\textwidth}   
            \centering 
\tikzset{%
    mylabel/.style={font=\normalsize},
    pics/entity/.style n args={3}{code={%
        \node[draw,
        rectangle split,
        rectangle split parts=2,
        text height=1ex,
        text width=9em,
        ] (#1)
        {#2 \nodepart[font=\small]{second}
            \begin{tabular}{>{\raggedright\arraybackslash}p{10em}}
                #3
            \end{tabular}
        };%
    }}
}
\resizebox{0.98\linewidth}{!}{
\begin{tikzpicture}

\draw[xstep=1,ystep=1,white,thin] (-1.25,-0.5) grid (4.75,8.5);
\draw (current bounding box.north east) -- (current bounding box.north west) -- (current bounding box.south west) -- (current bounding box.south east) -- cycle;

\draw (0,2) circle [radius=0.23] node(tb)  {tb};
\draw (1,2) circle [radius=0.23] node(tr)  {tr};
\draw (1.5,3) circle [radius=0.23] node(tp1) {tp};
\draw (1.5,4) circle [radius=0.23] node(nb1) {nb};
\draw (2.25,4) circle [radius=0.23] node(tp2) {tp};
\draw (3,4) circle [radius=0.23] node(nb2) {nb};
\draw (3.25,5) circle [radius=0.23] node(tp3) {tp};
\draw (4,5) circle [radius=0.23] node(nb3) {nb};


\draw (0.5, 3) node(bt1) {\LARGE$\leftouterjoin$};
\draw (0.5, 4) node(bt2) {\LARGE$\leftouterjoin$};
\draw (0.5, 5) node(bt3) {\LARGE$\leftouterjoin$};
\draw (0.5, 6) node(bt4) {\LARGE$\leftouterjoin$};
\draw (0.5, 7) node(bt5) {\LARGE$\leftouterjoin$};
\draw (0.5, 8) node(sel) {\LARGE$\pi$};

\draw (2.5, 6) node(bt6) {\LARGE$\leftouterjoin$};
\draw (2.5, 5) node(bt7) {\LARGE$\leftouterjoin$};

\draw[->] (tb)  to (bt1) node[yshift=-1.4em, xshift=-1.5em]{$\sigma$};
\draw[->] (tr)  to (bt1);
\draw[->] (bt1) to (bt2);
\draw[->] (tp1) to (bt2) node[yshift=-1.4em, xshift=2em]{$\sigma$};
\draw[->] (bt2) to (bt3);
\draw[->] (nb1) to (bt3);
\draw[->] (bt3) to (bt4);
\draw[->] (bt7) to (bt4);
\draw[->] (tp2) to (bt7) node[yshift=-1.4em, xshift=-1.5em]{$\sigma$};
\draw[->] (nb2) to (bt7);
\draw[->] (bt4) to (bt5);
\draw[->] (bt6) to (bt5);
\draw[->] (tp3) to (bt6) node[yshift=-1.4em, xshift=0em]{$\sigma$};
\draw[->] (nb3) to (bt6);
\draw[->] (bt5) to (sel);

\end{tikzpicture}
}
            \vspace{-1\baselineskip}
            \caption[]%
            {{\small Query \#7, baseline}}    
            \label{fig-q7-bl}
        \end{subfigure}
        \begin{subfigure}[t]{0.24\textwidth}   
            \centering 
\tikzset{%
    mylabel/.style={font=\normalsize},
    pics/entity/.style n args={3}{code={%
        \node[draw,
        rectangle split,
        rectangle split parts=2,
        text height=1ex,
        text width=9em,
        ] (#1)
        {#2 \nodepart[font=\small]{second}
            \begin{tabular}{>{\raggedright\arraybackslash}p{10em}}
                #3
            \end{tabular}
        };%
    }}
}
\resizebox{0.98\linewidth}{!}{
\begin{tikzpicture}

\draw[xstep=1,ystep=1,white,thin] (-1.25,-0.5) grid (4.75,8.5);
\draw (current bounding box.north east) -- (current bounding box.north west) -- (current bounding box.south west) -- (current bounding box.south east) -- cycle;

\draw (0,3) circle [radius=0.23] node(tg) {tg};
\draw (1,3) circle [radius=0.23] node(tb) {tb};

\draw (2,3) circle [radius=0.23] node(tc)  {tc};
\draw (3,3) circle [radius=0.23] node(nb1) {nb};
\draw (3.25,4) circle [radius=0.23] node(tdw) {\scriptsize tdw};
\draw (4,4) circle [radius=0.23] node(nb2) {nb};
\draw (1.5,6) circle [radius=0.23] node(nb3) {nb};

\draw (0.5, 4) node(bt1) {\LARGE$\leftouterjoin$};
\draw (0.5, 5) node(bt2) {\LARGE$\leftouterjoin$};
\draw (0.5, 6) node(bt3) {\LARGE$\leftouterjoin$};
\draw (0.5, 7) node(bt4) {\LARGE$\leftouterjoin$};
\draw (0.5, 8) node(sel) {\LARGE$\pi$};

\draw (2.5, 4) node(bt5) {\LARGE$\leftouterjoin$};
\draw (2.5, 5) node(bt6) {\LARGE$\leftouterjoin$};

\draw[->] (tg)  to (bt1);
\draw[->] (tb)  to (bt1) node[yshift=-1.4em, xshift=1.5em]{$\sigma$};
\draw[->] (bt1) to (bt2);
\draw[->] (tc) to  (bt5);
\draw[->] (nb1) to (bt5);
\draw[->] (bt5) to (bt2);
\draw[->] (tdw) to (bt6);
\draw[->] (nb2) to (bt6);
\draw[->] (bt2) to (bt3);
\draw[->] (bt6) to (bt3);
\draw[->] (bt3) to (bt4);
\draw[->] (nb3) to (bt4);
\draw[->] (bt4) to (sel);

\end{tikzpicture}
}
            \vspace{-1\baselineskip}
            \caption[]%
            {{\small Query \#7, 1NF}}    
            \label{fig-q7-1nf}
        \end{subfigure}
        \begin{subfigure}[t]{0.24\textwidth}   
            \centering 
\tikzset{%
    mylabel/.style={font=\normalsize},
    pics/entity/.style n args={3}{code={%
        \node[draw,
        rectangle split,
        rectangle split parts=2,
        text height=1ex,
        text width=9em,
        ] (#1)
        {#2 \nodepart[font=\small]{second}
            \begin{tabular}{>{\raggedright\arraybackslash}p{10em}}
                #3
            \end{tabular}
        };%
    }}
}
\resizebox{0.98\linewidth}{!}{
\begin{tikzpicture}

\draw[xstep=1,ystep=1,white,thin] (-0.5,-0.5) grid (5.5,8.5);
\draw (current bounding box.north east) -- (current bounding box.north west) -- (current bounding box.south west) -- (current bounding box.south east) -- cycle;

\draw (0,3) circle [radius=0.23] node(tb) {tb};
\draw (1,3) circle [radius=0.23] node(ta) {ta};

\draw (1.5,4) circle [radius=0.23] node(tg)  {tg};
\draw (3.25,5) circle [radius=0.23] node(nb1) {nb}; 
\draw (4.25,4) circle [radius=0.23] node(tdw)   {\scriptsize tdw};
\draw (5,4) circle [radius=0.23] node(nb2)   {nb};
\draw (3,4) circle [radius=0.23] node(nb3)   {nb};
\draw (2.25,4) circle [radius=0.23] node(tc)    {tc};

\draw (0.5, 4) node(bt1) {\LARGE$\leftouterjoin$};
\draw (0.5, 5) node(bt2) {\LARGE$\leftouterjoin$};
\draw (0.5, 6) node(bt0) {\LARGE$\leftouterjoin$};
\draw (0.5, 7) node(bt3) {\LARGE$\leftouterjoin$};
\draw (0.5, 8) node(sel) {\LARGE$\pi$};

\draw (2.5, 6) node(bt4) {\LARGE$\leftouterjoin$};
\draw (4.5, 5) node(bt5) {\LARGE$\leftouterjoin$};
\draw (2.5, 5) node(bt6) {\LARGE$\leftouterjoin$};

\draw[->] (tb)  to (bt1) node[yshift=-1.4em, xshift=-1.5em]{$\sigma$};
\draw[->] (ta)  to (bt1) node[yshift=-1.4em, xshift=1.5em]{$\sigma$};
\draw[->] (bt1) to (bt2);
\draw[->] (bt2) to (bt0);
\draw[->] (bt0) to (bt3);
\draw[->] (bt4) to (bt3);
\draw[->] (bt5) to (bt4);
\draw[->] (tg) to  (bt2);
\draw[->] (nb1) to (bt4);
\draw[->] (tdw) to (bt5);
\draw[->] (nb2) to (bt5);
\draw[->] (bt3) to (sel);

\draw[->] (nb3) to (bt6);
\draw[->] (tc) to  (bt6);
\draw[->] (bt6) to (bt0);
\end{tikzpicture}
}
            \vspace{-1\baselineskip}
            \caption[]%
            {{\small Query \#7, 2NF}}    
            \label{fig-q7-2nf}
        \end{subfigure}
        \begin{subfigure}[t]{0.24\textwidth}   
            \centering 
\tikzset{%
    mylabel/.style={font=\normalsize},
    pics/entity/.style n args={3}{code={%
        \node[draw,
        rectangle split,
        rectangle split parts=2,
        text height=1ex,
        text width=9em,
        ] (#1)
        {#2 \nodepart[font=\small]{second}
            \begin{tabular}{>{\raggedright\arraybackslash}p{10em}}
                #3
            \end{tabular}
        };%
    }}
}
\resizebox{0.98\linewidth}{!}{
\begin{tikzpicture}

\draw[xstep=1,ystep=1,white,thin] (-1.5,-0.5) grid (4.5,8.5);
\draw (current bounding box.north east) -- (current bounding box.north west) -- (current bounding box.south west) -- (current bounding box.south east) -- cycle;

\draw (0,0) circle [radius=0.23] node(tb) {tb};
\draw (1,0) circle [radius=0.23] node(tw) {tw};

\draw (1.5,1) circle [radius=0.23] node(nb1)  {nb};
\draw (1.5,2) circle [radius=0.23] node(ta)   {ta};
\draw (1.5,3) circle [radius=0.23] node(tg)   {tg};
\draw (1.5,4) circle [radius=0.23] node(td)   {td};
\draw (2.25,4) circle [radius=0.23] node(nb2)    {nb};
\draw (3,4) circle [radius=0.23] node(tc)     {tc};
\draw (1.5,6) circle [radius=0.23] node(nb3)  {nb};

\draw (0.5, 1) node(bt1) {\LARGE$\leftouterjoin$};
\draw (0.5, 2) node(bt2) {\LARGE$\leftouterjoin$};
\draw (0.5, 3) node(bt3) {\LARGE$\leftouterjoin$};
\draw (0.5, 4) node(bt4) {\LARGE$\leftouterjoin$};
\draw (0.5, 5) node(bt5) {\LARGE$\leftouterjoin$};
\draw (0.5, 6) node(bt6) {\LARGE$\leftouterjoin$};
\draw (0.5, 7) node(bt7) {\LARGE$\leftouterjoin$};
\draw (0.5, 8) node(sel) {\LARGE$\pi$};

\draw (2.5, 5) node(bt8) {\LARGE$\leftouterjoin$};

\draw[->] (tb)  to (bt1) node[yshift=-1.4em, xshift=-1.5em]{$\sigma$};
\draw[->] (tw)  to (bt1);
\draw[->] (bt1) to (bt2);
\draw[->] (bt2) to (bt3);
\draw[->] (bt3) to (bt4);
\draw[->] (bt4) to (bt5);
\draw[->] (bt5) to (bt6);
\draw[->] (bt6) to (bt7);
\draw[->] (bt8) to (bt6);
\draw[->] (nb1) to (bt2);
\draw[->] (ta) to  (bt3) node[yshift=-1.4em, xshift=2.25em]{$\sigma$};
\draw[->] (tg) to  (bt4);
\draw[->] (td) to  (bt5);
\draw[->] (nb2) to (bt8);
\draw[->] (tc) to  (bt8);
\draw[->] (nb3) to (bt7);
\draw[->] (bt7) to (sel);

\end{tikzpicture}
}
            \vspace{-1\baselineskip}
            \caption[]%
            {{\small Query \#7, 4NF}}
            \label{fig-q7-4nf}
        \end{subfigure}
    \caption{Query trees for queries 5..7; nodes represent tables, bowties joins, $\sigma$s selection and $\pi$s projection}
    \label{fig-query-trees-2}
\end{figure*}
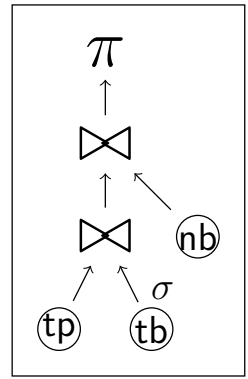
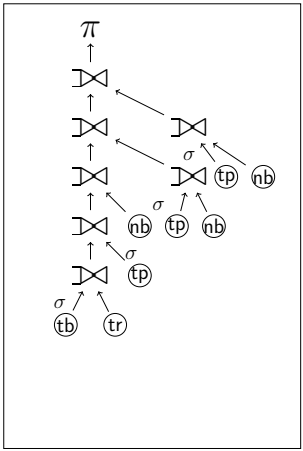


Figures~\ref{fig-query-trees-1} and \ref{fig-query-trees-2} show the complexity of the seven queries in the test suite for each of the four databases. With a few exceptions in query complexities between the baseline and 1NF database, the results show rather uniformly that query complexity tends to increase with further normalization. The query execution plans in the supplementary material are provided to describe the physical operations and query complexities explicitly -- the execution times in the query execution plans are inconsequential as they are dependent on hardware, and they are not executed in any controlled manner.

\subsection{Query performance}
\label{sec-results-query-performance}

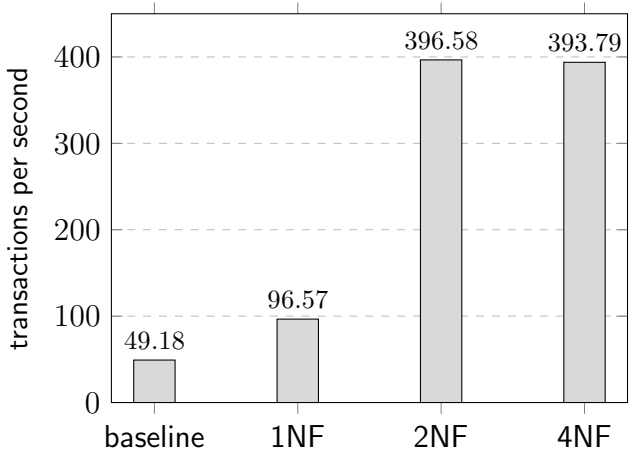
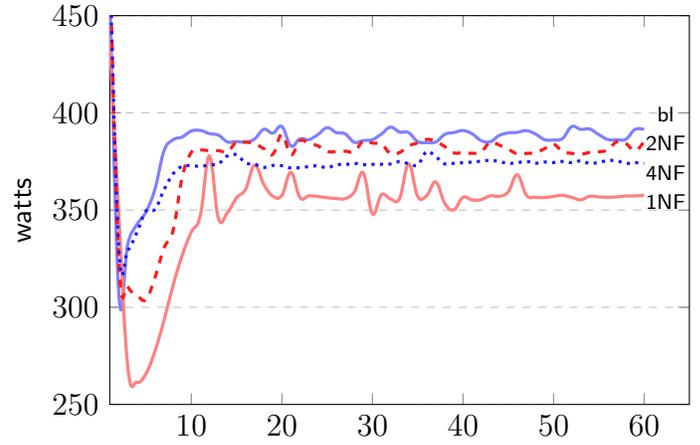
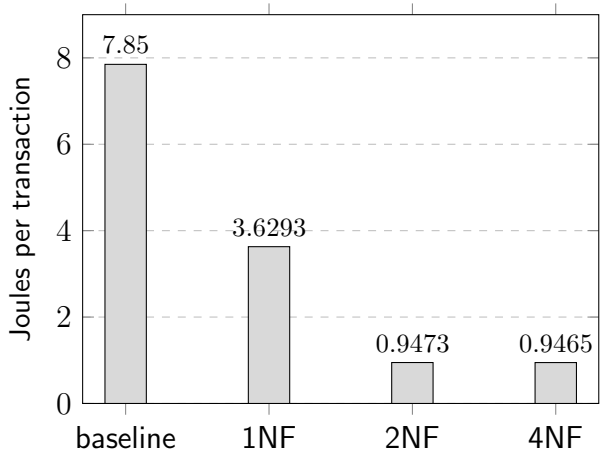
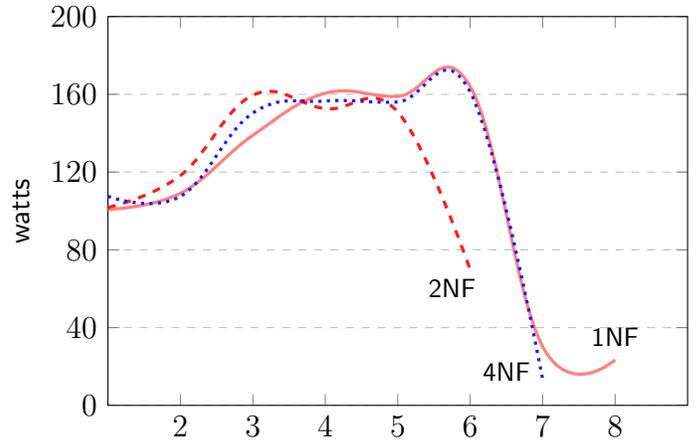
\begin{figure*}[h!]
    \centering
\centering
\begin{subfigure}[b]{0.48\textwidth}
        \pgfplotstableread[row sep=\\,col sep=&]{
            interval & tps     \\
            baseline & 49.181  \\
            1NF      & 96.571  \\
            2NF      & 396.581 \\
            4NF      & 393.790 \\
        }\mydata
\begin{tikzpicture}
    \begin{axis}[
    ybar,
    bar width=.55cm,
    width=1.0\textwidth,
    height=.80\textwidth,
    symbolic x coords={baseline, 1NF, 2NF, 4NF},
    xtick=data,
    ytick={0,100,200,300,400},
    nodes near coords,
    nodes near coords align={vertical},
    ymin=0,ymax=450,
    ylabel=\small transactions per second,
    ymajorgrids=true,
    grid style=dashed,
    legend pos=north east,
    every node near coord/.append style={font=\footnotesize},
    ]
    \addplot[style={draw=black,fill=light-gray}] table[x=interval,y=tps]{\mydata};
    \end{axis}
\end{tikzpicture}
\caption{Throughput averages of three test runs, measured in transactions per second (\textit{tps}) for each database}
\label{fig-throughput}
\end{subfigure}
\hfill
\begin{subfigure}[b]{0.48\textwidth}
\centering
\begin{tikzpicture}
    \begin{axis}[
    width=1.1\textwidth,
    height=.80\textwidth,
    ylabel={\small watts},
    xmin=1, xmax=65, 
    ymin=250, ymax=450,
    xtick={0,10,20,30,40,50,60},
    ytick={250,300,350,400,450},
    legend pos=north west,
    ymajorgrids=true,
    grid style=dashed,
    scaled y ticks=false, 
]     
    \addplot[color=blue!50,smooth,tension=0.7,very thick] table [x index=0,y index=1,col sep=space] {
a   b
1	466.9947222
2	306.70
3	332.86
4	342.54
5	349.02
6	360.91
7	380.52
8	386.14
9	387.60
10	390.64
11	390.96
12	389.35
13	388.52
14	385.06
15	385.23
16	384.96
17	386.60
18	391.54
19	389.57
20	392.9959444
21	383.00
22	386.55
23	386.27
24	388.89
25	392.36
26	389.88
27	389.19
28	384.87
29	385.77
30	386.1767222
31	389.15
32	392.63
33	390.14
34	389.83
35	385.24
36	384.70
37	385.03
38	388.95
39	389.87
40	390.9198889
41	389.26
42	385.76
43	385.12
44	388.35
45	389.67
46	389.97
47	390.86
48	389.14
49	386.24
50	386.0487222
51	386.82
52	392.90
53	391.52
54	391.64
55	388.80
56	386.29
57	386.09
58	386.46
59	391.49
60	391.4544253    
    };
    \addplot[color=red!50,smooth,tension=0.7,very thick] table [x index=0,y index=1,col sep=space] {
a   b
1	439.3056667
2	347.81
3	269.11
4	261.45
5	265.87
6	277.57
7	293.36
8	311.43
9	326.74
10	338.64
11	347.16
12	377.84
13	345.77
14	348.71
15	356.31
16	360.68
17	373.13
18	363.27
19	357.15
20	357.5101111
21	369.39
22	356.43
23	357.32
24	356.95
25	356.36
26	355.93
27	355.92
28	359.27
29	369.13
30	348.0916111
31	357.87
32	355.76
33	355.88
34	373.72
35	357.08
36	357.44
37	364.44
38	353.34
39	350.14
40	356.5413889
41	355.19
42	355.34
43	357.30
44	357.08
45	359.39
46	368.18
47	357.63
48	356.57
49	356.42
50	357.0513333
51	357.56
52	356.54
53	355.29
54	357.11
55	356.78
56	356.60
57	356.74
58	357.24
59	357.25
60	357.6635057
    };
    \addplot[color=red!90,smooth,tension=0.7,very thick,dashed] table [x index=0,y index=1,col sep=space] {
a   b
1	485.3123333
2	323.04
3	310.51
4	306.24
5	303.95
6	315.39
7	330.53
8	339.57
9	366.85
10	379.63
11	380.86
12	380.57
13	380.55
14	380.04
15	384.97
16	384.93
17	384.08
18	381.60
19	378.65
20	388.8498333
21	378.47
22	385.60
23	384.40
24	382.96
25	380.23
26	379.97
27	380.27
28	380.14
29	385.49
30	384.4373333
31	383.91
32	379.10
33	379.61
34	382.97
35	384.55
36	386.34
37	385.38
38	382.60
39	379.41
40	379.5594444
41	379.19
42	379.86
43	384.90
44	383.44
45	381.27
46	378.92
47	379.15
48	379.54
49	381.53
50	383.8877778
51	380.06
52	379.19
53	378.72
54	379.63
55	380.01
56	380.98
57	385.20
58	384.14
59	380.53
60	385.2871264
};
    \addplot[color=blue!90,smooth,tension=0.7,very thick,dotted] table [x index=0,y index=1,col sep=space] {
a   b
1	481.1946667
2	330.16
3	329.43
4	338.62
5	349.50
6	350.44
7	360.51
8	368.63
9	372.50
10	372.82
11	372.48
12	373.25
13	373.80
14	377.88
15	378.41
16	374.27
17	373.53
18	371.96
19	372.96
20	372.8237778
21	371.64
22	371.82
23	372.37
24	372.75
25	371.77
26	373.29
27	373.41
28	374.01
29	373.34
30	373.4986111
31	373.56
32	373.69
33	374.90
34	373.87
35	373.25
36	380.04
37	377.42
38	373.75
39	374.07
40	374.3577222
41	374.68
42	375.14
43	375.57
44	374.99
45	373.98
46	374.42
47	374.95
48	374.11
49	374.71
50	374.4257778
51	375.01
52	374.40
53	375.27
54	373.97
55	374.52
56	375.48
57	375.28
58	374.02
59	374.43
60	374
    };
    \node [above] at (614, 140) {\scriptsize bl};
    \node [above] at (614, 95)  {\scriptsize 1NF};
    \node [above] at (614, 110) {\scriptsize 4NF};
    \node [above] at (614, 125) {\scriptsize 2NF};
    
    \end{axis}
\end{tikzpicture}

    \caption{Power consumption averages of three test runs for each database over 60 minutes; baseline database abbreviated as ``bl''}
    \label{fig-power_consumption}
\end{subfigure}
\vspace{+1\baselineskip}

\begin{subfigure}[t]{0.48\textwidth}  
    \centering 
        \pgfplotstableread[row sep=\\,col sep=&]{
            interval & indices  \\
            baseline & 7.85 \\
            1NF      & 3.6293 \\
            2NF      & 0.9473 \\
            4NF      & 0.9465 \\
        }\mydata
    \pgfplotsset{scaled y ticks=false}
\begin{tikzpicture}
    \begin{axis}[
    ybar,
    bar width=.55cm,
    width=1.0\textwidth,
    height=.80\textwidth,
    scaled y ticks=false, 
    y tick label style={/pgf/number format/fixed},
    symbolic x coords={baseline, 1NF, 2NF, 4NF},
    xtick=data,
    nodes near coords,
    ymin=0,ymax=9,
    scaled ticks=false,
    ylabel=\small Joules per transaction,
    ymajorgrids=true,
    grid style=dashed,
    every node near coord/.append style={font=\footnotesize,/pgf/number format/precision=4},
    ]
    \addplot[style={draw=black,fill=light-gray}] table[,x=interval,y=indices]{\mydata};  
    \end{axis}
\end{tikzpicture}
    \caption{Energy consumption averages of three test runs per transaction for each database}  
    \label{fig-ec_per_transaction}
\end{subfigure}
\hfill
\begin{subfigure}[t]{0.48\textwidth}
\centering
\begin{tikzpicture}
    \begin{axis}[
    width=1.1\textwidth,
    height=.80\textwidth,
    ylabel={\small watts},
    xmin=1, xmax=9,
    ymin=0, ymax=200,
    xtick={2,3,4,5,6,7,8},
    ytick={0,40,80,120,160,200},
    legend pos=north west,
    ymajorgrids=true,
    grid style=dashed,
    scaled y ticks=false, 
]     
    \addplot[color=red!50,smooth,tension=0.7,very thick] table [x index=0,y index=1,col sep=space] {
a   b
1   100.755
2   109.095
3   138.945
4   160.54
5   158.945
6   163.99
7   30.225
8   23.4
    };
    \addplot[color=red!90,smooth,tension=0.7,very thick,dashed] table [x index=0,y index=1,col sep=space] {
a   b
1   101.7315
2   118.21
3   159.54
4   152.625
5   150.915
6   69.795
    };
    \addplot[color=blue!90,smooth,tension=0.7,very thick,dotted] table [x index=0,y index=1,col sep=space] {
a   b
1   107.51
2   107.515
3   150.385
4   156.57
5   156.095
6   161.195
7   13.425
    };

    \node [above] at (700, 25) {\footnotesize 1NF};
    \node [above] at (475, 50) {\footnotesize 2NF};
    \node [above] at (550, 7)  {\footnotesize 4NF};
    
    \end{axis}
\end{tikzpicture}
    \caption{Power consumption for database creation over time (minutes); shorter lines indicate shorter creation times}
    \label{fig-power_con_creation}
\end{subfigure}
    \caption{Throughput averages show the 2NF and 4NF databases outperforming the 1NF database; transaction processing in the 1NF database is subject to more fluctuation than transactions processed in the 2NF and 4NF databases; on average, the transactions processed in the 2NF and 4NF databases consume less energy than transactions processed in the 1NF database;  the creation of the 2NF database consumes less power and time when compared to the 4NF database, which in turn consumes less power and time than the creation of the 1NF database;}
    \label{fig-query-performance}
\end{figure*}

\texttt{pgbench} reports average throughput. The averages of the three tests runs for each database showed consistent results, implying that there was, e.g., no interference from background processes. Throughput in the baseline schema ranged from 49.074 transactions per second (\textit{tps}) to 49.393, in 1NF schema from 96.557 to 96.590, in 2NF schema from 392.545 to 399.452, and in 4NF schema from 393.790 to 395.726, showing that the largest fluctuation in \textit{tps} within one database was less than 2\% over three one-hour test runs. A summary of throughput is illustrated in Fig.~\ref{fig-throughput}.

\subsection{Power and energy consumption}
\label{sec-results-energy-consumption}

Power consumption for each database in the transaction processing tests is illustrated in Fig.~\ref{fig-power_consumption}. In Fig.~\ref{fig-power_consumption} it can be seen that after approximately 15 minutes, the power consumptions stabilize. The overall energy consumptions are approximately on 385.366 Wh (baseline database, \textit{SD} = 38.40), 350.578 Wh (1NF database, \textit{SD} = 32.00), 375.954 Wh (2NF database, \textit{SD} = 43.84), and 372.760 Wh (4NF database, \textit{SD} = 23.25), while serving different number of transactions for the used wattage, shown in Joules in Fig.~\ref{fig-ec_per_transaction}.


Power consumption for database creation, including data insertion and creation of all indices is shown in Fig.~\ref{fig-power_con_creation}. Shorter lines in the graph indicate that the database was created in less amount of time. Total energy consumptions were approximately 52.4 KJ (1NF), 43.3 KJ (2NF), and 51.0 KJ (4NF). 



\section{Discussion}
\label{sec-discussion}

\subsection{A summary of the results}

In summary, normalization from 1NF to either 2NF or 4NF increased database efficiency in terms of smaller database size on disk, larger throughput despite increase in query complexity, and smaller energy consumption per transaction. However, the effects of normalization from 2NF to 4NF were negligible. Additionally, the 1NF database creation, while in bulk rather than naturally over time, took 31\% more time than the creation of the 2NF database, and 16\% more time than the creation of the 4NF database. Given the size of the databases and that the creation time is measured in minutes, in IMDb's case the differences in creation time are not particularly impactful.

A rather interesting result is that the throughput in the 4NF database is significantly larger than in the 1NF database, and that the difference between throughputs of the 4NF and 2NF are negligible, despite the fact that six of the seven queries contain more joins in the 4NF database than in the 1NF database, and three of the seven queries contain more joins in the 4NF database than in the 2NF database. Despite more joins, which in theory imply slower query execution time and smaller throughput, the results may be explained by the fact that the average number of rows per table is smallest in the 4NF database, and consequently, joining smaller tables is less expensive in this case than creating temporary tables at runtime from large tables, even while indexed. Additionally, some operations are inherently bound to different hardware components, such as joins to memory, sorting to CPU, and sequential scans retrieving a larger number of rows to I/O.

Some of the fluctuation in \textit{tps} (Fig.~\ref{fig-throughput}) can probably be attributed to \texttt{pgbench}, which picks one of the seven queries in the test suite at random, when creating a new client with a request for PostgreSQL. As some of the queries are more complex (i.e., potentially slower to execute) than others, test runs where slower queries were picked more often than faster ones arguably result in lower \textit{tps}. The power fluctuations in Fig.~\ref{fig-power_con_creation} are most likely explained by task switching from the data insertion to index creation.

\subsection{Practical implications}
\label{sec-disc-practical-impl}

The results support the common notion that 1NF is not a sufficient level of normalization in transaction processing, and the results yielded by this study show that normalization from 1NF to 2NF reduces storage space by approximately 10\% while increasing throughput by approximately 311\% and reducing energy consumption by approximately 74\%.

Running a single-node database server of this size at full capacity for one hour consumes roughly the same amount of power as driving an electric vehicle for 0.6 to 1.0 miles (1.0 to 1.6 km), depending on the vehicle\footnote{https://ev-database.org/imp/cheatsheet/energy-consumption-electric-car}. It is worth remembering that these numbers cover only the database server, not aspects such as hardware cooling on infrastructure level, network, software applications, or any client devices. In the case of this study, the level of normalization determines how many concurrent clients can be served with this energy consumption.

Both energy consumption (i.e., how many clients one can serve with the hardware) and storage space requirements (i.e., how much hardware is needed to store the database) contribute to the green data considerations. Additionally, more transaction processing and storage hardware naturally increase the costs of the whole system. Finally, energy may be produced with different means, and from a holistic perspective, simply considering energy consumption without accounting for the source of energy is hardly the whole picture.

This study was not an attempt to provide generalizable results. In fact, we are sceptical on the generalizability of the results, given all the considerations discussed in the next Section and Section~\ref{sec-bg-efficiency}. Rather, we hope to provide at least some evidence which can be used in tandem with other similar studies on different datasets, configurations, and business domains to further understand how well results such as these \textit{can generalize}. Further, this study was not an attempt to argue that the ``IMDb database should be normalized to normal form \textit{x}''. Rather, the IMDb dataset was used as a vehicle for attempting to determine some effects of database normalization using a real-world dataset, and the baseline database results are included for illustrative purposes rather than for fair comparison. We emphasize, again, that the results yielded by this study should not be interpreted as ``normalization from 1NF to 2NF will always reduce energy consumption by 74\%''. Rather, the results show that normalization indeed reduces energy consumption with the IMDb database running on the hardware specified in this study, with seven read-only operations with a certain isolation level, and with a single-node PostgreSQL instance. Based on this study, we certainly agree that fair performance testing is both challenging as well as arduous when the potential lack of generalizability is considered.


\subsection{Limitations and threats to validity}
\label{sec-disc-threats}

We recognize the pitfalls described in prior studies on performance testing. To that end, we try to both follow the \textit{Fair Benchmarking} \cite{Raasveldt_2018} checklist and justify our approach when we will not or can not, as well as transparently report the technical details behind our research setting. It is worth noting that some of the checklist items are intented for performance comparisons between two different DBMSs, and these do not apply to our research setting. The checklist and our adherence to it is summarized in Table~\ref{table-checklist}. Unfortunately, the checklist is not always particularly verbose regarding its items.

\textit{The queries}: First, described in Section~\ref{sec-setting-optimization}, the seven SQL statements run by \texttt{pgbench} are all read operations due to the nature of the IMDb dataset, which does not include the most write-heavy data objects such as user data or reviews. This limits our results, as the test suite does not account for transaction contention and subsequent locks. However, the real-world IMDb database is arguably read-heavy overall. Second, our queries do not necessarily reflect the real queries the IMDb system handles. This is partly due to the fact that the publicly available IMDb dataset is incomplete, but also due to the fact that our queries are merely \textit{speculated} queries based on the IMDb website. For this reason, our queries do not ``cover the whole evaluation space'' \cite{Raasveldt_2018}, but only seven read-only statements, which are arguably at least based on queries that have possible functionality. Third, corner cases are not tested: arguably at any given point in time there are certain titles which draw more inquiries than others from the end-users. This potentially leads to situations in which these titles are already ``closer to the user'' at the time of querying. That is, as different end-users query similar titles, the data are potentially already in caches and memory, consequently reducing latency. In our tests, the titles were randomly selected, and given the number of titles and the number of transactions executed, it is unlikely (although not impossible) that a title that was recently queried is queried again. As such a scenario is unlikely, it seems reasonable to argue that such rare occurrences had any noticeable effects on the results of this study. Additionally, the queried data object such as title or person is chosen randomly, and while this randomization is done artificially with SQL due to limits of \texttt{pgbench}, this is similar to all databases and the performance implications are minimal. Fourth, \texttt{pgbench} randomly executes one of the seven SQL queries on each run, effectively meaning that each of the seven queries is executed approximately the same number of times. This does not necessarily reflect real-world use-cases, but since we had no insights on ratio of different queries, we chose not to speculate, and treated the queries equally. In the case the ratios of queries is different, the theoretical implications on the results depends on the commonness of the query. For example, if query \#7 is more common than others, the 4NF database would arguably perform worse relative to the other databases. On the other hand, if query \#2 was more common, the differences between the database performances would possibly be smaller. All in all, a goal more important than to mimic real-world use-cases as accurately as possible is to provide a setting to compare the different databases as fairly as possible.

\begin{table*}
  \small
  \centering
  \caption{How this study adheres to the \textit{Fair Benchmarking} \cite{Raasveldt_2018} checklist; in cases we did not adhere to the checklist, please see the corresponding sections where the reasons behind these choices are explained}
  \label{table-checklist}
  \begin{tabular}{ll}
\toprule
Fair Benchmarking checklist item & Approach in this study \\
\midrule
Benchmark covers whole evaluation space (ES)
& \textcolor{BrickRed}{\faTimesCircle} We cover only common queries (Section~\ref{sec-setting-optimization}).
\\
Justify picking benchmark subset 
& \textcolor{OliveGreen}{\faCheckCircle} Based on IMDb use-cases (Section~\ref{sec-setting-normalization}).
\\
Benchmark stresses functionality in the ES
& \textcolor{OliveGreen}{\faCheckCircle} Based on IMDb use-cases (Section~\ref{sec-setting-normalization}).
\\
{[Report]} hardware configuration 
& \textcolor{OliveGreen}{\faCheckCircle} cf. Table~\ref{table-hardware}.
\\
{[Report]} DBMS parameters and version 
& \textcolor{OliveGreen}{\faCheckCircle} cf. Supplementary material (\texttt{config}) and Table~\ref{table-hardware}.
\\
{[Report]} source code or binary files 
& \textcolor{OliveGreen}{\faCheckCircle} PostgreSQL source code is publicly available.
\\
{[Report]} data, schema \& queries 
& \textcolor{OliveGreen}{\faCheckCircle} cf. Fig.~\ref{fig-schemas} and Supplementary material (\texttt{inputs}).
\\
{[Report]} compilation flags 
& \textcolor{Dandelion}{\faExclamationCircle} Not applicable. We did not compile from source.
\\
{[Report]} system parameters 
& \textcolor{OliveGreen}{\faCheckCircle} cf. Table~\ref{table-hardware}.
\\
Similar functionality 
& \textcolor{OliveGreen}{\faCheckCircle} All databases were subject to similar queries.
\\
Equivalent workload 
& \textcolor{OliveGreen}{\faCheckCircle} The same benchmark was used for all databases.
\\
{[Utilize]} different data 
& \textcolor{BrickRed}{\faTimesCircle} We used only the IMDb dataset (Section~\ref{sec-disc-threats}).
\\
{[Utilize]} various workloads 
& \textcolor{BrickRed}{\faTimesCircle} We tested only read operations (Section~\ref{sec-setting-optimization}).
\\
Differentiate between cold and hot runs 
& \textcolor{OliveGreen}{\faCheckCircle} All databases were tested starting with cold runs.
\\
Cold runs: Flush OS and CPU caches 
& \textcolor{OliveGreen}{\faCheckCircle} Cache flush and hard reboots (Section~\ref{sec-setting-performance}).
\\
Hot runs: Ignore initial runs 
& \textcolor{OliveGreen}{\faCheckCircle} Comparable runs with \texttt{pgbench}.
\\
Ensure preprocessing is the same between systems 
& \textcolor{OliveGreen}{\faCheckCircle} All tests were run in the same environment.
\\
Be aware of automatic index creation 
& \textcolor{OliveGreen}{\faCheckCircle} All indices were created manually (Section~\ref{sec-setting-size}).
\\
Verify results 
& \textcolor{OliveGreen}{\faCheckCircle} cf. Sections \ref{sec-results-query-performance} and \ref{sec-results-energy-consumption}.
\\
Test different datasets 
& \textcolor{BrickRed}{\faTimesCircle} We used only the IMDb dataset (Section~\ref{sec-disc-threats}).
\\
Corner cases work 
& \textcolor{Dandelion}{\faExclamationCircle} Not applicable. Corner cases are unknown.
\\
Do several runs to reduce interference 
& \textcolor{OliveGreen}{\faCheckCircle} Read-operation tests were run multiple times.
\\
Check standard deviation for multiple runs 
& \textcolor{OliveGreen}{\faCheckCircle} cf. Sections \ref{sec-results-query-performance} and \ref{sec-results-energy-consumption}.
\\
Report robust metrics 
& \textcolor{OliveGreen}{\faCheckCircle} cf. Sections \ref{sec-results-query-performance} and \ref{sec-results-energy-consumption}.
\\
\bottomrule
\end{tabular}
\end{table*}

\textit{The dataset}: We tested the effects of normalization with the IMDb dataset. ``Test different data sets'' \cite{Raasveldt_2018} is probably targeted for benchmarks in which the data is computer-generated (as opposed to real-world data), and we saw no need to benchmark computer-generated datasets, as in this case, we focus particularly on the IMDb database. As the dataset is real-world data for the part that is available, we did not feel the need to consider other datasets, as would arguably be the case if the data were synthetic. The dataset, as stated before, is not the complete IMDb database. Beyond data structures such as those concerning users and title reviews, which are evident by browsing the IMDb website, it is unclear how complex the entire database is.

\textit{Database design}: Beyond normalization, none of the \textit{logical} database structures are optimized. For example, creating a derived table which contains the top-250 titles which are updated periodically rather than calculated every time query \#1 is run would arguably reduce latency significantly, while resulting only in a small storage requirement increase. On the other hand, the \textit{physical} database structures are optimized for the seven queries through indices.

\textit{PostgreSQL settings and hardware}: All the tests were run with the same PostgreSQL configuration. While we tried to control the effects of different configurations prior to the tests proper, different configurations did not yield any particular performance gains for any of the databases. However, these tests were not exhaustive, and it should be noted that each of the databases could likely be optimized further through configuring PostgreSQL. Additionally, hardware presents some challenges to fair testing, both in DBMS-DBMS comparison and in our research setting. Comparing performance with different hardware seems hardly fair, yet some hardware favors some environments over others. Some operations are CPU, some I/O, and some memory bound, and as, e.g., the number of joins in queries grow, more CPU time is usually needed. However, in these cases, less memory may be needed as the number of rows per table is smaller. For replication purposes, the configuration file should be adjusted to match hardware. 

Furthermore, as we tested read-only queries, this study does not account for possible data anomalies arising from redundant data. The potential of such problems is in theory reduced through normalizing the database further. There is also a possibility of human error in several aspects of the research setting. In conclusion, we agree that measuring performance should be transparent, and the results of such studies interpreted in their respective contexts rather than as generalizable to all contexts. 

\subsection{Future research avenues}
\label{sec-disc-future}

While important, data system efficiency considerations should not be limited to the metrics measured in this study. For example, while there were differences in energy consumption between the database schemas, it is arguably important to acknowledge the energy sources, some of which are less green than others. On the other hand, efforts towards understanding efficiency in a holistic manner need to acknowledge the effects of software development environments on efficiency. For example, it is recognized that many software libraries and object-relational mappers have deficiencies in producing optimized queries \cite{Liu_2023}. While these performance problems can be fixed, it takes time, which is arguably an important facet of efficiency. Furthermore, studies have shown concern on the effects of DBMS compiler error messages on query formulation, especially among novices \cite{Taipalus_2023_framework}. In summary, we propound the view towards considering efficiency as a whole. While this is challenging in both research and industry, it should be acknowledged that the efforts towards efficient data systems should be accounted as well. That is, a data system that is optimized beyond diminishing returns may in practice be less efficient than a system that is ``good enough'', when all optimization efforts are taken into account. 

It has been noted before that performance comparison studies do not necessarily generalize due to large number of different variables in different environments \cite{Taipalus_2024_perf}, and that the comparisons should be implemented and reported transparently. In our opinion, the potential lack of generalizability should not be interpreted in a way that performance comparisons should not be studied, but rather that database performance needs a high number of specific studies to be understood. Additionally, the performance comparison in this study was founded on a read-only transaction processing database. It has been shown that data models in, e.g., machine learning applications have different considerations \cite{Wan_2019}, which puts forward the need to study performance in different settings as well. 

\section{Conclusion}
\label{sec-concl}

The theoretical effects of database normalization on performance have been rather commonly accepted, although their magnitude has not received ample scientific attention. In this study, we compared efficiency of the IMDb dataset normalized to different normal forms using PostgreSQL and read-only operations. The results show that normalization from 1NF to 2NF not only save storage space, but also significantly increases throughput and significantly decreases energy consumption per transaction. These results are applicable in logical relational database design, especially in guiding the decisions towards higher normal forms.

\section*{Statements and Declarations}

The authors have no competing interests to declare that are relevant to the content of this article.

The raw data reported in this study and replication package for the results are publicly available at 
\href{https://anonymous.4open.science/r/f4-normalization-74E7}{https://anonymous.4open.science/r/f4-normalization-74E7}.

 \bibliographystyle{elsarticle-num} 
 \bibliography{cas-refs}





\end{document}